\documentclass[twocolumn,trackchanges,twocolappendix]{aastex63}

\newcommand{\msun}{\mathrm{M}_\odot}

\defcitealias{1991A&A...248..485D}{DM91}
\defcitealias{1992ApJ...396..178F}{FM92}
\defcitealias{2017ApJS..230...15M}{MS17}
\defcitealias{2018AJ....156..257S}{S18}

\received{Dec 3, 2025}
\revised{Jan 23, 2026}
\accepted{Feb 3, 2026}

\usepackage{graphicx}
\usepackage[T1]{fontenc}
\usepackage{aecompl}
\usepackage[]{amsmath}

\usepackage{color}

\usepackage{xspace}

\usepackage{soul}
\begin{document}

\title{The Binary Fraction of Stars in the Dwarf Galaxy Ursa Minor via Dark Energy Spectroscopic Instrument}

\author[0000-0002-4900-2088]{Tian Qiu}\thanks{qtskyxt@sjtu.edu.cn}
\affiliation{Department of Astronomy, School of Physics and Astronomy, and Key Laboratory for Particle Astrophysics and Cosmology (MOE)/Shanghai Key Laboratory for Particle Physics and Cosmology, Shanghai Jiao Tong University, Shanghai 200240, People's Republic of China}
\affiliation{State Key Laboratory of Dark Matter Physics, School of Physics and Astronomy, Shanghai Jiao Tong University, Shanghai 200240, China}
\author[0000-0002-5762-7571]{Wenting Wang}\thanks{wenting.wang@sjtu.edu.cn}
\affiliation{Department of Astronomy, School of Physics and Astronomy, and Key Laboratory for Particle Astrophysics and Cosmology (MOE)/Shanghai Key Laboratory for Particle Physics and Cosmology, Shanghai Jiao Tong University, Shanghai 200240, People's Republic of China}
\affiliation{State Key Laboratory of Dark Matter Physics, School of Physics and Astronomy, Shanghai Jiao Tong University, Shanghai 200240, China}
\author[0000-0003-2644-135X]{Sergey E. Koposov}\thanks{Sergey.Koposov@ed.ac.uk}
\affiliation{Institute for Astronomy, University of Edinburgh, Royal Observatory, Blackford Hill, Edinburgh EH9 3HJ, UK}
\affiliation{Institute of Astronomy, University of Cambridge, Madingley Road, Cambridge CB3 0HA, UK}
\author[0000-0002-9110-6163]{Ting S. Li}
\affiliation{Department of Astronomy \& Astrophysics, University of Toronto, Toronto, ON M5S 3H4, Canada}
\author[0000-0002-7393-3595]{Nathan R. Sandford}
\affiliation{Department of Astronomy \& Astrophysics, University of Toronto, Toronto, ON M5S 3H4, Canada}
\author[0000-0002-5758-150X]{Joan Najita}
\affiliation{NSF NOIRLab, 950 N. Cherry Ave., Tucson, AZ 85719, USA}
\author[0000-0002-6469-8263]{Songting Li}
\affiliation{Department of Astronomy, School of Physics and Astronomy, and Key Laboratory for Particle Astrophysics and Cosmology (MOE)/Shanghai Key Laboratory for Particle Physics and Cosmology, Shanghai Jiao Tong University, Shanghai 200240, People's Republic of China}
\affiliation{Tsung-Dao Lee Institute, Shanghai Jiao Tong University, Shanghai, 201210, China}
\affiliation{State Key Laboratory of Dark Matter Physics, School of Physics and Astronomy, Shanghai Jiao Tong University, Shanghai 200240, China}
\author[0000-0002-8010-6715]{Jiaxin Han}
\affiliation{Department of Astronomy, School of Physics and Astronomy, and Key Laboratory for Particle Astrophysics and Cosmology (MOE)/Shanghai Key Laboratory for Particle Physics and Cosmology, Shanghai Jiao Tong University, Shanghai 200240, People's Republic of China}
\affiliation{State Key Laboratory of Dark Matter Physics, School of Physics and Astronomy, Shanghai Jiao Tong University, Shanghai 200240, China}
\author[0000-0002-4928-4003]{Arjun Dey}
\affiliation{NSF NOIRLab, 950 N. Cherry Ave., Tucson, AZ 85719, USA}
\author[0000-0002-6667-7028]{Constance Rockosi}
\affiliation{Department of Astronomy and Astrophysics, UCO/Lick Observatory, University of California, 1156 High Street, Santa Cruz, CA 95064, USA}
\affiliation{Department of Astronomy and Astrophysics, University of California, Santa Cruz, 1156 High Street, Santa Cruz, CA 95065, USA}
\affiliation{University of California Observatories, 1156 High Street, Sana Cruz, CA 95065, USA}
\author[0000-0002-2761-3005]{Boris Gaensicke}
\affiliation{Department of Physics, University of Warwick, Gibbet Hill Road, Coventry, CV4 7AL, UK}
\author[0000-0002-6800-5778]{Jesse Han}
\affiliation{Center for Astrophysics $|$ Harvard \& Smithsonian, 60 Garden Street, Cambridge, MA 02138, USA}
\author{Benjamin Alan Weaver}
\affiliation{NSF NOIRLab, 950 N. Cherry Ave., Tucson, AZ 85719, USA}
\author{Adam D. Myers}
\affiliation{Department of Physics \& Astronomy, University of Wyoming, 1000 E. University, Dept.~3905, Laramie, WY 82071, USA}
\author{Jessica Nicole Aguilar}
\affiliation{Lawrence Berkeley National Laboratory, 1 Cyclotron Road, Berkeley, CA 94720, USA}
\author[0000-0001-6098-7247]{Steven Ahlen}
\affiliation{Department of Physics, Boston University, 590 Commonwealth Avenue, Boston, MA 02215 USA}
\author[0000-0002-0084-572X]{Carlos Allende Prieto}
\affiliation{Instituto de Astrof\'{\i}sica de Canarias, C/ V\'{\i}a L\'{a}ctea, s/n, E-38205 La Laguna, Tenerife, Spain}
\affiliation{Departamento de Astrof\'{\i}sica, Universidad de La Laguna (ULL), E-38206, La Laguna, Tenerife, Spain}
\author[0000-0001-9712-0006]{Davide Bianchi}
\affiliation{Dipartimento di Fisica ``Aldo Pontremoli'', Universit\`a degli Studi di Milano, Via Celoria 16, I-20133 Milano, Italy}
\affiliation{INAF-Osservatorio Astronomico di Brera, Via Brera 28, 20122 Milano, Italy}
\author{David Brooks}
\affiliation{Department of Physics \& Astronomy, University College London, Gower Street, London, WC1E 6BT, UK}
\author{Todd Claybaugh}
\affiliation{Lawrence Berkeley National Laboratory, 1 Cyclotron Road, Berkeley, CA 94720, USA}
\author[0000-0002-1769-1640]{Axel de la Macorra}
\affiliation{Instituto de F\'{\i}sica, Universidad Nacional Aut\'{o}noma de M\'{e}xico, Circuito de la Investigaci\'{o}n Cient\'{\i}fica, Ciudad Universitaria, Cd. de M\'{e}xico C.~P.~04510, M\'{e}xico}
\author{Peter Doel}
\affiliation{Department of Physics \& Astronomy, University College London, Gower Street, London, WC1E 6BT, UK}
\author[0000-0002-3033-7312]{Andreu Font-Ribera}
\affiliation{Institut de F\'{i}sica d’Altes Energies (IFAE), The Barcelona Institute of Science and Technology, Edifici Cn, Campus UAB, 08193, Bellaterra (Barcelona), Spain}
\author[0000-0002-2890-3725]{Jaime E. Forero-Romero}
\affiliation{Departamento de F\'isica, Universidad de los Andes, Cra. 1 No. 18A-10, Edificio Ip, CP 111711, Bogot\'a, Colombia}
\affiliation{Observatorio Astron\'omico, Universidad de los Andes, Cra. 1 No. 18A-10, Edificio H, CP 111711 Bogot\'a, Colombia}
\author[0000-0001-9632-0815]{Enrique Gaztañaga}
\affiliation{Institut d'Estudis Espacials de Catalunya (IEEC), c/ Esteve Terradas 1, Edifici RDIT, Campus PMT-UPC, 08860 Castelldefels, Spain}
\affiliation{Institute of Cosmology and Gravitation, University of Portsmouth, Dennis Sciama Building, Portsmouth, PO1 3FX, UK}
\author[0000-0003-3142-233X]{Satya Gontcho A Gontcho}
\affiliation{Lawrence Berkeley National Laboratory, 1 Cyclotron Road, Berkeley, CA 94720, USA}
\affiliation{University of Virginia, Department of Astronomy, Charlottesville, VA 22904, USA}
\author{Gaston Gutierrez}
\affiliation{Fermi National Accelerator Laboratory, PO Box 500, Batavia, IL 60510, USA}
\author[0000-0001-8528-3473]{Jorge Jimenez}
\affiliation{Institut de F\'{i}sica d’Altes Energies (IFAE), The Barcelona Institute of Science and Technology, Edifici Cn, Campus UAB, 08193, Bellaterra (Barcelona), Spain}
\author[0000-0003-0201-5241]{Dick Joyce}
\affiliation{NSF NOIRLab, 950 N. Cherry Ave., Tucson, AZ 85719, USA}
\author[0000-0003-3510-7134]{Theodore Kisner}
\affiliation{Lawrence Berkeley National Laboratory, 1 Cyclotron Road, Berkeley, CA 94720, USA}
\author[0000-0002-6731-9329]{Claire Lamman}
\affiliation{The Ohio State University, Columbus, 43210 OH, USA}
\author[0000-0003-1838-8528]{Martin Landriau}
\affiliation{Lawrence Berkeley National Laboratory, 1 Cyclotron Road, Berkeley, CA 94720, USA}
\author[0000-0001-7178-8868]{Laurent Le Guillou}
\affiliation{Sorbonne Universit\'{e}, CNRS/IN2P3, Laboratoire de Physique Nucl\'{e}aire et de Hautes Energies (LPNHE), FR-75005 Paris, France}
\author[0000-0002-1125-7384]{Aaron Meisner}
\affiliation{NSF NOIRLab, 950 N. Cherry Ave., Tucson, AZ 85719, USA}
\author{Ramon Miquel}
\affiliation{Institut de F\'{i}sica d’Altes Energies (IFAE), The Barcelona Institute of Science and Technology, Edifici Cn, Campus UAB, 08193, Bellaterra (Barcelona), Spain}
\affiliation{Instituci\'{o} Catalana de Recerca i Estudis Avan\c{c}ats, Passeig de Llu\'{\i}s Companys, 23, 08010 Barcelona, Spain}
\author[0000-0001-9070-3102]{Seshadri Nadathur}
\affiliation{Institute of Cosmology and Gravitation, University of Portsmouth, Dennis Sciama Building, Portsmouth, PO1 3FX, UK}
\author[0000-0002-0644-5727]{Will Percival}
\affiliation{Department of Physics and Astronomy, University of Waterloo, 200 University Ave W, Waterloo, ON N2L 3G1, Canada}
\affiliation{Perimeter Institute for Theoretical Physics, 31 Caroline St. North, Waterloo, ON N2L 2Y5, Canada}
\affiliation{Waterloo Centre for Astrophysics, University of Waterloo, 200 University Ave W, Waterloo, ON N2L 3G1, Canada}
\author{Claire Poppett}
\affiliation{University of California, Berkeley, 110 Sproul Hall \#5800 Berkeley, CA 94720, USA}
\affiliation{Space Sciences Laboratory, University of California, Berkeley, 7 Gauss Way, Berkeley, CA 94720, USA}
\affiliation{Lawrence Berkeley National Laboratory, 1 Cyclotron Road, Berkeley, CA 94720, USA}
\author[0000-0001-7145-8674]{Francisco Prada}
\affiliation{Instituto de Astrof\'{i}sica de Andaluc\'{i}a (CSIC), Glorieta de la Astronom\'{i}a, s/n, E-18008 Granada, Spain}
\author[0000-0001-6979-0125]{Ignasi Pérez-Ràfols}
\affiliation{Departament de F\'isica, EEBE, Universitat Polit\`ecnica de Catalunya, c/Eduard Maristany 10, 08930 Barcelona, Spain}
\author{Graziano Rossi}
\affiliation{Department of Physics and Astronomy, Sejong University, 209 Neungdong-ro, Gwangjin-gu, Seoul 05006, Republic of Korea}
\author[0000-0002-9646-8198]{Eusebio Sanchez}
\affiliation{CIEMAT, Avenida Complutense 40, E-28040 Madrid, Spain}
\author{David Schlegel}
\affiliation{Lawrence Berkeley National Laboratory, 1 Cyclotron Road, Berkeley, CA 94720, USA}
\author[0000-0002-3461-0320]{Joseph Harry Silber}
\affiliation{Lawrence Berkeley National Laboratory, 1 Cyclotron Road, Berkeley, CA 94720, USA}
\author{David Sprayberry}
\affiliation{NSF NOIRLab, 950 N. Cherry Ave., Tucson, AZ 85719, USA}
\author[0000-0003-1704-0781]{Gregory Tarlé}
\affiliation{University of Michigan, 500 S. State Street, Ann Arbor, MI 48109, USA}
\author[0000-0001-5381-4372]{Rongpu Zhou}
\affiliation{Lawrence Berkeley National Laboratory, 1 Cyclotron Road, Berkeley, CA 94720, USA}
\author[0000-0002-6684-3997]{Hu Zou}
\affiliation{National Astronomical Observatories, Chinese Academy of Sciences, A20 Datun Road, Chaoyang District, Beijing, 100101, P.~R.~China}



\begin{abstract}

We utilize multi-epoch line-of-sight velocity measurements from the Milky Way Survey of the Dark Energy Spectroscopic Instrument to estimate the binary fraction for member stars in the dwarf spheroidal galaxy Ursa Minor. 
Our dataset comprises 670 distinct member stars, with a total of more than 2,000 observations collected over approximately one year.
We constrain the binary fraction for UMi to be $0.61^{+0.16}_{-0.20}$ and $0.69^{+0.19}_{-0.17}$, with the binary orbital parameter distributions based on solar neighborhood observation from Duquennoy \& Mayor (1991) and Moe \& Di Stefano (2017), respectively. Furthermore, by dividing our data into two subsamples at the median metallicity, we identify that the binary fraction for the metal-rich ([Fe/H]>-2.14) population is slightly higher than that of the metal-poor ([Fe/H]<-2.14) population. Based on the Moe \& Di Stefano model, the best-constrained binary fractions for metal-rich and metal-poor populations in UMi are $0.86^{+0.14}_{-0.24}$ and $0.48^{+0.26}_{-0.19}$, respectively.
After a thorough examination, we find that this offset cannot be attributed to sample selection effects. We also divide our data into two subsamples according to their projected radius to the center of UMi, and find that the more centrally concentrated population in a denser environment has a lower binary fraction of $0.33^{+0.30}_{-0.20}$, compared with $1.00^{+0.00}_{-0.32}$ for the subsample in more outskirts.

\end{abstract}



\section{Introduction} 
\label{sec:intro}

Binary stars play a crucial role in our understanding of the star formation process, galactic dynamics, and the evolution history of galaxies \citep{2001ApJ...556..265W,2013ARA&A..51..269D, 2018ApJ...854..147B}.
Binaries are common in our Milky Way Galaxy, with studies showing that binary systems are identified more often than single stars in the solar neighborhood \citep{1969JRASC..63..275H,1991A&A...248..485D,2010ApJS..190....1R}.
These earlier studies mostly focus on the statistical properties of binary stars within a limited range, typically extending only to about 30 pc from the Sun.

With advances in observational techniques, investigations of binary populations have extended \citep{2018AJ....156...18P} beyond the solar neighborhood to open clusters \citep{2002AJ....123.1570P,2020ApJ...901...49L}, globular clusters \citep{1988AJ.....96..123P,1992PASP..104..981H,2007MNRAS.380..781S,2012A&A...540A..16M} and dwarf galaxies \citep{1996AJ....111..750O,2010ApJ...722L.209M,2013ApJ...779..116M}. 
In dense stellar environments, binaries are likely to have undergone significant dynamical processes such as stellar encounters \citep{2011A&A...528A.144K,2020MNRAS.496.5176D}, which can disrupt binary systems. Compared with globular clusters, dwarf galaxies are more diffuse, making close stellar encounters among their member stars less frequent. The lower stellar densities in dwarf galaxies may allow a higher fraction of binaries to survive than those in dense globular clusters.

Most of the classical dwarf galaxies around our Milky Way are distant, and thus binaries in dwarf galaxies are difficult to resolve directly. The influence of binaries is often inferred from their contribution to observed line-of-sight velocity (LOSV) variations. For example, binary orbital motions can inflate the measured velocity dispersion of a galaxy, thereby complicating dynamical mass estimates
\citep{1996AJ....111..750O,1996MNRAS.279..108H,2010ApJ...722L.209M,2011ApJ...736..146K,2019MNRAS.487.2961M,2023ApJ...956...91W}.
With advances in spectroscopic surveys, the number of stars observed in dwarf galaxies has increased substantially, from less than a hundred in earlier studies \citep{1995A&A...300...31Q,1995AJ....110.2131A,1995AJ....109..151V,2002MNRAS.330..792K,2003ApJ...588L..21K} to several hundred or even over a thousand in those known dwarf galaxies such as Carina, Fornax, Sculptor, Sextans, etc \citep{2009AJ....137.3100W, 2023ApJS..268...19W}. 

The methodological foundation was established by these early pioneering efforts to search for individual binary systems.
\citet{1988AJ.....96..123P} employed multi-epoch radial-velocity data of giants to identify binary candidates,
while \citet{1992PASP..104..981H} provided a comprehensive review of different binary-detection techniques and their applications in dense stellar systems. 
Later works extended their efforts beyond individual binary identification to the development of statistical approaches for constraining the overall binary fraction.

In particular,
\citet{1996AJ....111..750O} analyzed extensive radial velocities to search binaries and infer their frequency,
and demonstrated that the binaries can significantly affect the measured velocity dispersion in dense systems.
Based on this points, \citet{2010ApJ...721.1142M} introduced a method to estimate the binary fraction in dwarf galaxies based on their velocity dispersion. \citet{2011ApJ...733...46S} applied this approach to Segue 1 to determine the velocity dispersion, removing the influence from the binaries. A slightly later study by \citet{2013ApJ...779..116M} conducted a detailed binary analysis of several dwarf galaxies, including Carina, Fornax, etc. Combining several previous datasets, \citet{2017ApJ...836..202S} compiled multi-epoch LOSV measurements and applied a newly developed method to study the binary population in Leo II. A more recent study by \citet{2018AJ....156..257S} (hereafter \citetalias{2018AJ....156..257S}) further extended this methodology by combining LOSV measurements from multiple sources collected over approximately 30 years, allowing for an investigation of the binary fractions in Draco and Ursa Minor. Most recently, \citet{2023A&A...677A..95A} applied this approach to measure the binary fraction in the Sculptor dwarf galaxy. 

These cumulative efforts highlight the growing interest in dwarf galaxies as valuable targets for binary population studies. 
The inferred binary fractions vary significantly among different dwarf galaxies, indicating that binary populations are potentially sensitive to the specific conditions of their host systems.
Dwarf galaxies themselves exhibit a wide range of structural, chemical, and dynamical properties \citep[e.g.][]{2012AJ....144....4M,2024arXiv241107424P}. Though dwarf galaxies are mostly old, the star formation histories of different dwarf galaxies still vary \citep{2014ApJ...789..147W}. Variations in stellar density, metallicity, and star formation histories among different dwarf galaxies provide a natural laboratory for investigating how these factors influence the formation, evolution, and long-term survival of binary stars.

The Ursa Minor (UMi) dwarf spheroidal galaxy is one of the most well-known satellites of the Milky Way, located at a distance of approximately 76 kpc \citep{1995AJ....110.2131A, 2002AJ....123.3199C}. 
UMi is characterized by an old stellar population with a mean age exceeding 10 Gyr, making it an important target for understanding early galactic evolution and chemical enrichment \citep{2001ApJ...549L..63M,2010ApJS..191..352K}.
In recent years, numerous spectroscopic observations of UMi have been conducted, yielding a wealth of LOSV measurements for a large sample of stars in UMi \citep{2003ApJ...588L..21K,2004ApJ...611L..21W,2010ApJS..191..352K}, which offer a valuable opportunity to probe the binary populations in UMi via the velocity variations.

In this study, we take advantage of the multi-epoch LOSV measurements from the Dark Energy Spectroscopic Instrument (DESI) to estimate the binary fraction in UMi. 
DESI is one of the most advanced multi-object spectrographs designed for wide-field surveys \citep{2013arXiv1308.0847L,2016arXiv161100036D, 2016arXiv161100037D, 2022AJ....164..207D,2023AJ....166..259S}.
The DESI survey aims to observe more than 40 million galaxies, including targets from the Bright Galaxy Survey \citep[BGS; ][]{2023AJ....165..253H}, Luminous Red Galaxies \citep[LRG; ][]{2023AJ....165...58Z}, Emission Line Galaxies \citep[ELG; ][]{2023AJ....165..126R} and quasars \citep{2023ApJ...944..107C} to explore the nature of dark energy \citep{2023AJ....165...50M,2023ApJ...943...68L,2023AJ....165..124A,2024AJ....167...62D}. Significant cosmological results from DESI's Early Data Release \citep[EDR; ][]{2024AJ....168...58D}, Data Release 1 \citep[DR1; ][]{2025arXiv250314745D} and the future Data Release 2 (DR2) include two-point clustering statistics \citep{2024arXiv241112020D}, baryon acoustic oscillations (BAO) signals \citep{2024arXiv240403000D,2025JCAP...01..124A, 2025PhRvD.112h3514A, 2025PhRvD.112h3515A}, galaxy clustering analyses \citep{2024arXiv241112021D}, and updated cosmological constraints \citep{2025JCAP...02..021A,2024arXiv241112022D}.


Alongside the cosmological surveys of DESI conducted at dark time with good observing conditions, the DESI Milky Way Survey (MWS) \citep{2023ApJ...947...37C} takes advantage of observing conditions suboptimal for galaxy redshift measurements, such as poor seeing and bright time with significant moonlight contamination. The primary DESI MWS program focuses on stars 
at Galactic latitudes $|b| > 20^\circ$, reaching up to a magnitude limit of $r = 19$. More than seven million stars are selected from the DESI Legacy Imaging Survey \citep{2019AJ....157..168D} for observation under this program. Its footprint 
encompasses numerous streams, satellite galaxies, and globular clusters in the Milky Way. So far, the stellar value-added catalogs have been published based on the DESI EDR \citep{2024MNRAS.533.1012K} and DESI DR1 \citep{2025arXiv250514787K}, containing over 4 million unique stars with LOSV and stellar parameter measurements. Additionally, the DESI MWS includes secondary targets, such as faint blue horizontal branch (BHB) stars, complementing those observed in the main survey. Beyond these main and secondary targets in the main survey, the DESI MWS also employs ``tertiary'' dedicated tile programs \citep{2023AJ....165...50M} specifically to observe the M31 galaxy, as well as stellar streams and dwarf satellite galaxies within our Galaxy, including UMi. The unprecedented depth and high-precision measurements provided by DESI allow us to apply this methodology to a significantly larger sample of member stars in UMi. This also enables us to explore potential differences in binary fractions across various stellar populations within UMi.

This paper is structured as follows: In Section~\ref{sec:data}, we describe the properties of the DESI data used in this work. 
Section~\ref{sec:method} introduces our methodology for constructing the binary simulation and the likelihood function to estimate the binary fraction of a dwarf galaxy. 
In Section~\ref{sec:results}, we present the results for the dwarf galaxy UMi and the offset between the metal-rich and metal-poor populations. 
We discuss the potential causes for this binary fraction discrepancy in Section~\ref{sec:discussion}. 
Finally, the summary and conclusions are provided in Section~\ref{sec:conclusion}.

\section{Data}
\label{sec:data}

In this section, we describe the observational data utilized in this work.

\subsection{DESI}

DESI is designed to conduct sky surveys covering 14,000 deg$^2$ on the Mayall 4-meter telescope at the Kitt Peak National Observatory, Arizona \citep{2013arXiv1308.0847L,2016arXiv161100036D, 2016arXiv161100037D, 2022AJ....164..207D,2023AJ....166..259S}. 
It is a multi-object spectroscopic instrument capable of collecting up to 5,000 spectra per exposure, providing a continuous wavelength coverage from 3,600 to 9,824~\AA~ with a full width at half maximum resolution of about 1.8~\AA~ \citep{2023AJ....165....9S,2024AJ....168...95M,2024AJ....168..245P}.

\begin{figure}
\plotone{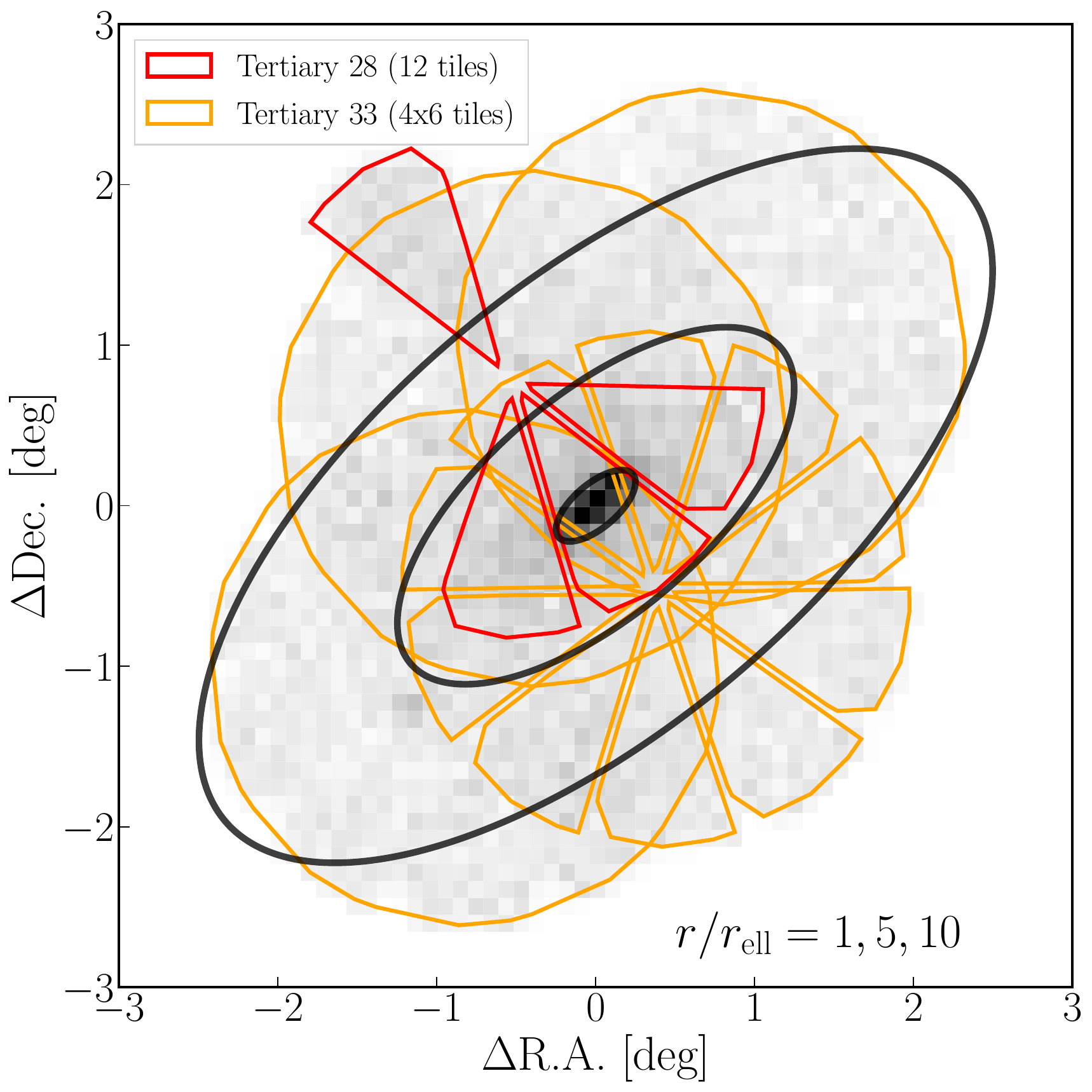}
\caption{Spatial distribution of stars observed by DESI tertiary programs in UMi, shown in a gnomonic projection. The footprints of Tertiary 28 and Tertiary 33 are illustrated by the red and orange regions, respectively. Black ellipses denote the 1, 5, and 10 elliptical half-light radii of UMi.
}
\label{fig:footprint}
\end{figure}

In this work, we utilize data collected through DESI MWS tertiary dedicated tile programs 28 and 33. Tertiary 28 observations were conducted on 2023 April 27 while only four of the ten DESI spectrograph petals were in operation. These observations include 12 tiles (i.e., unique fiber configurations) designed such that the inner $\simeq$5 half-light radii of UMi were covered by three adjacent working petals. Tertiary 33 observations were conducted on 7 separate nights between 2023 June 06 and 2024 March 27 while DESI was fully functional. These observations include 24 tiles distributed to 4 pointings such that each tile observes the inner half-light radius of UMi and the full program footprint extended out to $\simeq$10 half-light radii. The footprints of both programs are presented in Figure \ref{fig:footprint}. The adopted parameters of UMi used in this work are listed in Table \ref{tab:umi}. In both programs, each tile was observed between 350 and 1500 s depending on the observing conditions in order to achieve an effective exposure time of $\simeq$300 s (see \citealt{2023AJ....166..259S} for a description of how the effective exposure time is calculated). A more detailed description of these UMi tertiary observations will be presented in Sanford et al. (In Prep).

\begin{table}[h!]
\caption{Summary of adopted parameters of UMi.}
 \label{tab:umi}
\centering
 \begin{tabular}{|l l|} 
 \hline
 Distance & 76 kpc\\
 \citep{2002AJ....123.3199C}&\\
 \hline
 Half-light radii & 18.3 arcmin \\
 \citep{Pace2025OJAp....8E.142P}&\\
 \hline
 Ellipticity & 0.55 \\
 \citep{Pace2025OJAp....8E.142P}&\\
 \hline
 Position angle & 50 deg \\
 \citep{Pace2025OJAp....8E.142P}&\\
 \hline
  Velocity dispersion & 8.0 km~s$^{-1}$ \\
  (\citetalias{2018AJ....156..257S}) &\\
 \hline
 Average age for member stars & 10 Gyr \\
 \citep{2002AJ....123.3199C} &\\
 \hline
 Central luminosity density & 0.006 $L_\odot \mathrm{pc}^{-3}$ \\
 \citep{1998ARAA..36..435M} &\\
 \hline
 \end{tabular}
\end{table}

These observations were processed with the latest version of the DESI spectroscopic pipeline\citep{2023AJ....165..144G}, yielding 49,836 spectra for 8,460 unique objects. These stellar spectra provide valuable measurements, including LOSVs ($v_{\rm{los}}$), metallicity ([Fe/H]), surface gravity ($\log g $), effective temperature ($T_\mathrm{eff}$), a few other chemical abundances, etc., which are output by the MWS \textsc{rvs}\footnote{\url{https://github.com/segasai/rvspecfit}}\citep{2019ascl.soft07013K} and \textsc{sp} pipelines. Readers can refer to \cite{2023ApJ...947...37C} for more details about the MWS pipelines. This data will be included as a part of DESI DR2 scheduled to be released in spring 2027.

It is important to note that the uncertainties of LOSV ($\sigma_{v_{\rm{los}}}$) derived from the pipeline have a systematic floor of around 1-2~km~s$^{-1}$,
which refers to an irreducible component of measurement uncertainty that persists even when statistical errors are minimized.
It typically arises from instrumental effects such as spectrograph flexure or wavelength calibration errors and observing conditions such as lunation fraction, atmosphere, etc.
This floor is modeled as a constant added in quadrature to the formal uncertainty and represents a lower limit to the achievable precision.
As discussed in Section~4.3.2 of \citet{2025arXiv250514787K}, the systematic error floors are 1.0, 1.6, and 2.0~km~s$^{-1}$ for the MWS bright, dark, and backup programs \citep[see][for details about the MWS backup program]{2025arXiv250517230D}, respectively. Given the specific strategy of the tertiary program, we will additionally constrain the systematic error floor for our case to improve the precision of our results in Section~\ref{ssec:syserr}.

\subsection{Member Stars in UMi}
\label{sec:umimem}
\begin{figure}
\plotone{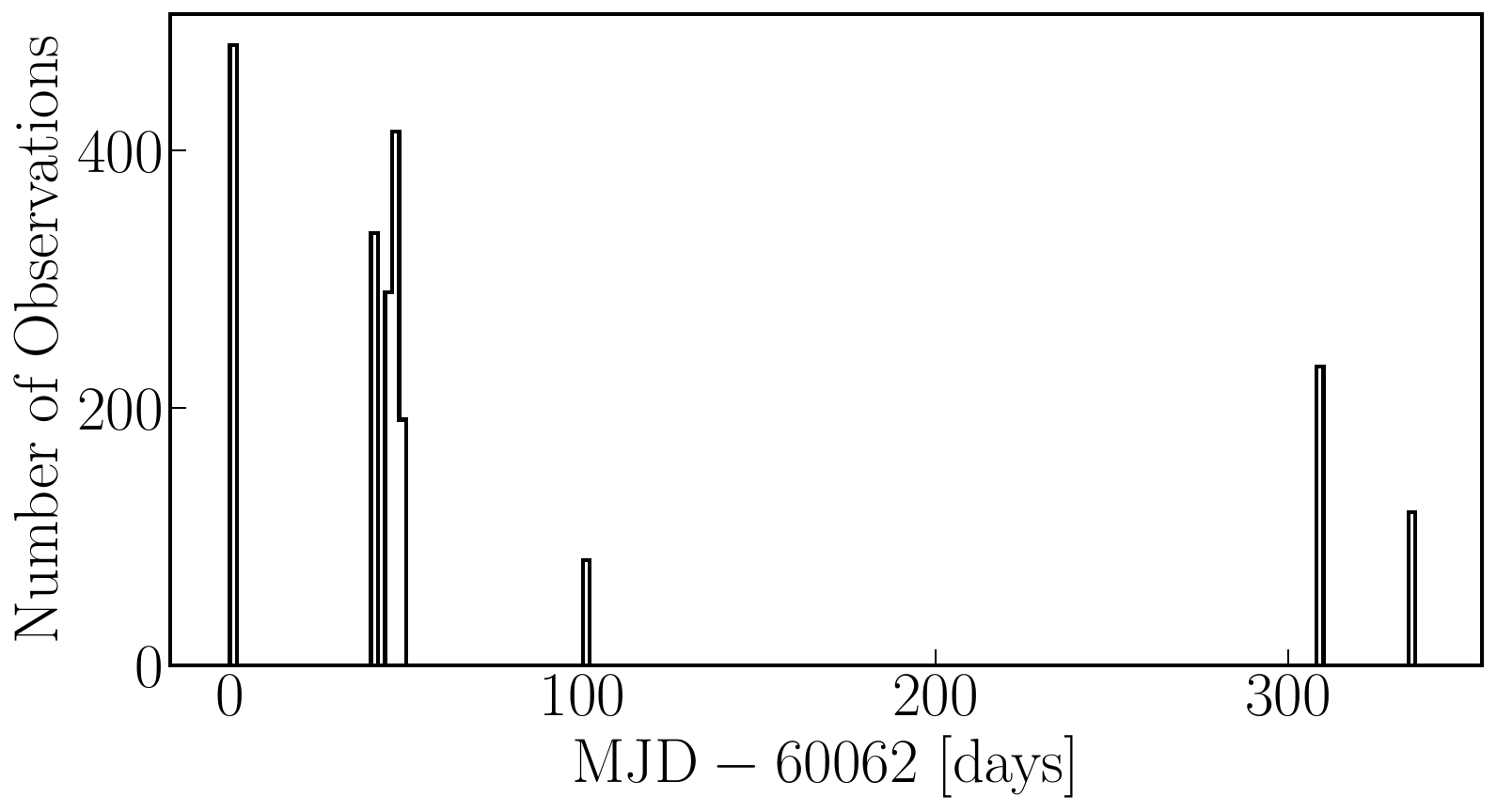}
\plotone{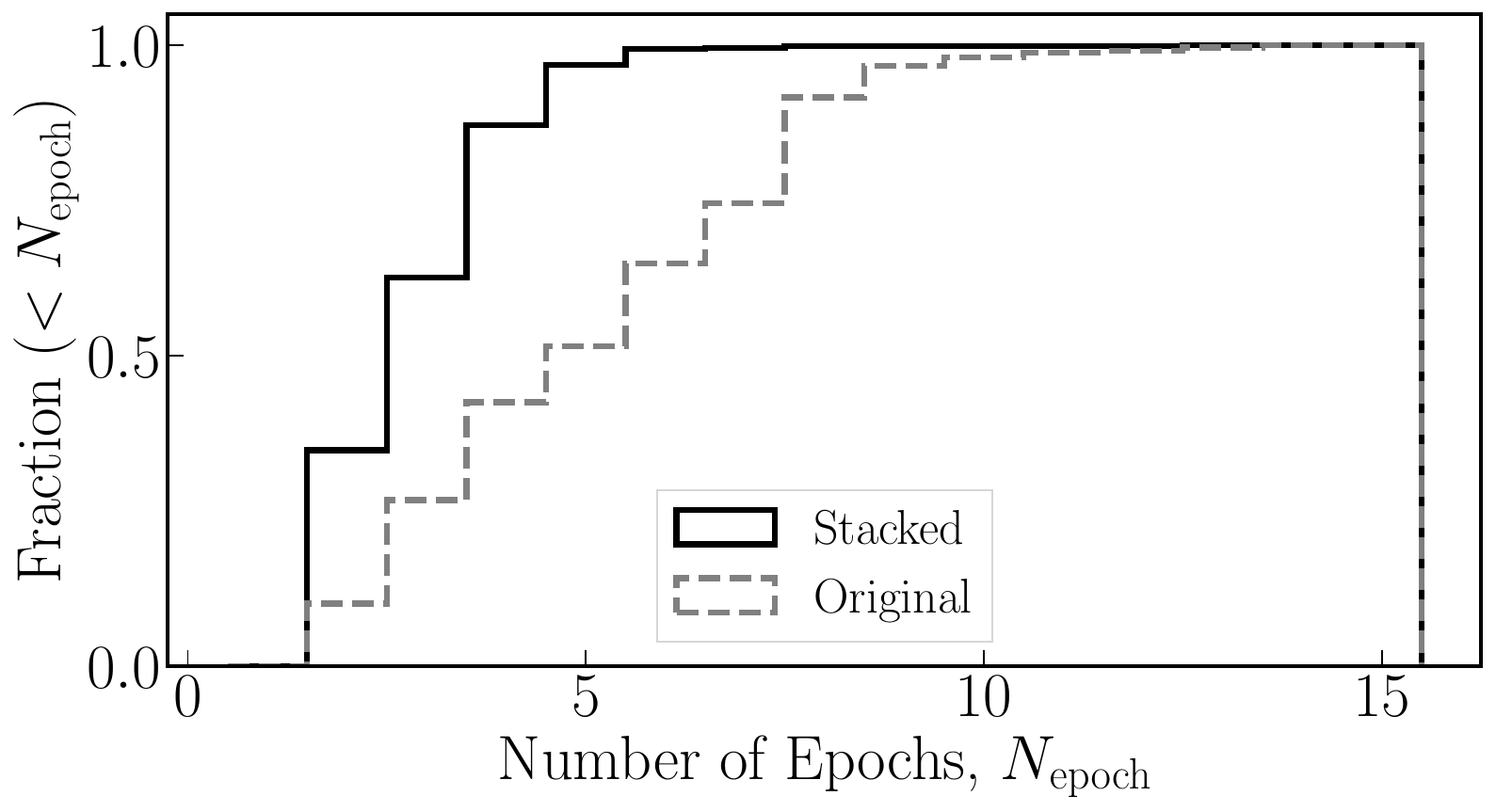}
\caption{The basic information of our selected member stars (670 in total) of UMi.
{\bf Top:} The histogram showing the MJD distribution for each epoch. Due to the observation strategy of the Tertiary Program, the MJD range for each star spans from several days to almost a year. 
{\bf Bottom:} Cumulative histograms of the number of epochs per star. The lighter dashed histogram shows the distribution before stacking multiple spectra taken on the same night (see Step 4 in Section~\ref{sec:umimem}), while the solid black histogram shows the distribution after merging those exposures into a single epoch.
\label{fig:time_info}}
\end{figure}

To obtain a clean sample of member stars in UMi, we match the stars observed by DESI in the UMi footprint with the highly probable (probability greater than 90\%) member star lists for dwarf spheroidal galaxies selected by \citet{2022ApJ...940..136P}\footnote{The catalog is available at \doi{10.5281/zenodo.6533295} \citep{andrew_b_pace_2025_14732612}}.
The authors combine astrometry from {\it Gaia} early data release 3 (EDR3) and the photometry from either DECam, {\it Gaia}, or Pan-Starrs to effectively separate the Milky Way foreground stars from member stars in these dwarf galaxies \citep{2019ApJ...875...77P,2020AJ....160..124M}.
They model the proper motions and spatial positions of stars around a satellite with Gaussian mixture models, with different Gaussian components standing for true member stars and foreground/background components. In particular, they assume a projected Plummer stellar distribution \citep{1911MNRAS..71..460P} for the spatial term.
After matching our data to their catalog, we obtain a sample of 1,721 candidates within the UMi footprint, comprising a total of 8,439 observations.
Each member star has multiple observations by DESI at different epochs over the course of a year.
The distributions of the Modified Julian Date (MJD) range and the number of epochs per star are shown in Figure~\ref{fig:time_info}.
For each member star, we adopt the metallicity and surface gravity values derived from the stacked spectra across all exposures and by the MWS \textsc{rvs} pipeline. 
These values serve as representative parameters for each individual object and will be used in the subsequent selection criteria.
Besides, for multiple observations obtained within a one-day interval, we treat them as one single measurement. Specifically, we co-add the original spectra taken on the same night and derive the LOSV and its corresponding uncertainty from the stacked spectrum rather than from individual exposures.

\begin{figure}
\plotone{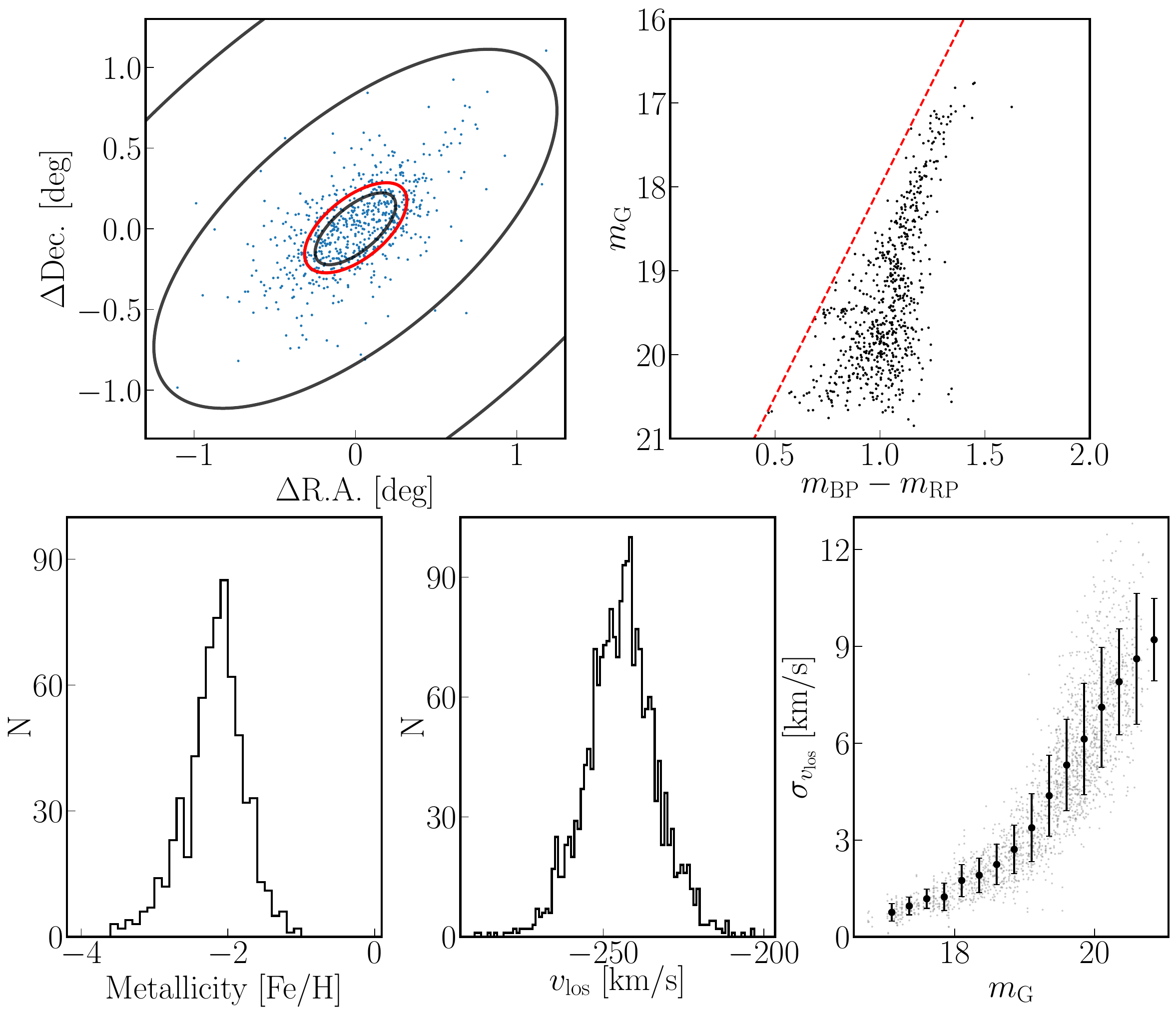}
\caption{{\bf Top-left panel}: 
Spatial distribution of UMi member stars in equatorial coordinates in a gnomonic projection. The red ellipse marks the division between the inner and outer subsamples based on projected radius from the galactic center (see Section~\ref{sec:discussion}). The three black ellipses correspond to the 1, 5, and 10 elliptical half-light radii, identical to those shown in Figure~\ref{fig:footprint}.
{\bf Top-right panel}: Color–magnitude diagram for UMi member stars. The red dashed line marks the color–magnitude selection used to exclude horizontal branch stars.
{\bf Bottom-left panel}: Metallicity distribution of UMi member stars, measured from stacked spectra combining all available exposures for each star.
{\bf Bottom-middle panel}: LOSV distribution for all measurements of the member stars, derived from stacked spectra obtained each night.
{\bf Bottom-right panel}: Magnitude dependence of LOSV uncertainties. Gray points denote individual stars, while black points with errorbars indicate the mean LOSV uncertainty and the corresponding standard error derived in magnitude bins of width 0.25.
}
\label{fig:info}
\end{figure}

To meet our purpose of properly determining the binary fraction in UMi, we perform some additional selections and preparations to the list of member stars:
\begin{enumerate}
    \item We cut the surface gravity at $\log g<4.0$. Since UMi is more than 60 kpc away, we can mainly observe giant stars, including red giants and blue horizontal branch stars. Therefore, we introduce a surface gravity cut to have a clean sample of giants.

    \item Stars with exposures all within one day are excluded from the analysis. Our binary fraction estimation relies on detecting variations in LOSV, which requires sufficient time baselines. In this study, we also model the Roche lobe (see Section~\ref{sec:method}), which effectively excludes binary systems with orbital periods shorter than 4 days. As a result, significant LOSV variations due to binary orbital motions are unlikely to occur within time spans shorter than a quarter of the orbital period, further justifying this exclusion criterion.
    
    \item We only use stars with LOSV uncertainties ($\sigma_{v_{\mathrm{los}}}$) smaller than 20~km~s$^{-1}$. Large uncertainties introduce noise and provide little insight into the LOSV variabilities due to binary orbital motions, which will be discussed later.
    
\end{enumerate}
Note that we exclude the entire star that does not meet criteria 1-2 above, whereas we remove individual observations rather than the entire star for criteria 3 above.
We do not apply any LOSV range cuts to avoid removing true binary systems with large velocity variations by mistake. Instead, we manually exclude six outlier stars that are confirmed not to be binaries.

In addition, we also match our data with the variable catalog from {\it Gaia} DR3 to remove contamination by RR Lyrae. However, there might be some faint RR Lyrae left, which are not identified by {\it Gaia}. Thus, we further exclude horizontal branch stars using an empirical cut of $m_G>23-5\times (m_{\rm BP}-m_{\rm RP})$.
After careful selection above, we have 670 unique member stars spanning over one day, comprising a total of 2,147 individual measurements of LOSV. We summarize these numbers in Table~\ref{tab:num}.
The distributions of basic information about our sample are presented in Figures~\ref{fig:time_info} and \ref{fig:info}. As can be seen from the upper left panel, most of our member stars are within 1-1.5 degrees (semi-major axis) to the center, i.e., within 5 times the elliptical half-light radius of UMi (see the middle black ellipse in Figure~\ref{fig:footprint}).

\begin{table}[h!]
\caption{Summary of UMi stars and their corresponding observations in the DESI  Tertiary Program.}
 \label{tab:num}
\centering
 \begin{tabular}{|l c c|} 
 \hline
  Category & Stars & Observations \\[0.5ex]
 \hline\hline 
 Total number in UMi footprint & 8,460 & 49,836 \\ 
 \hline
 Matched with \citet{2022ApJ...940..136P} & 1,721 & 8,439  \\
 \hline
 Final sample used in this work & 670 & 2,147  \\
 \hline
 \end{tabular}
\end{table}

\section{Methodology}
\label{sec:method}

Detecting binary stars in dwarf galaxies through astrometric methods poses significant challenges due to their substantial distances from us. 
The angular resolution required to resolve binaries at such distances is extremely high, making it feasible only for wide binaries with large separations. Even then, it is still rare to discover such systems. 
Consequently, most binary systems in dwarf galaxies remain unresolved in astrometric data. To address this limitation, we constrain the binary fraction in UMi by analyzing the LOSV variability of its member stars.

Motivated by the method developed in \citet{2017AJ....153..254S} and \citetalias{2018AJ....156..257S}, we introduce several improvements to achieve a more realistic constraint on the binary fraction and a more robust estimation of its uncertainty. 
Specifically, we refine both the modeling of orbital parameters and the construction of the likelihood function. The detailed methodological differences will be described in the following subsections.

This method does not attempt to identify individual binary systems. 
Instead, it treats all member stars as a collective sample and statistically examines the distribution of their LOSV variations. 
By comparing the observed LOSV variability with that of simulated samples corresponding to different assumed binary fractions, we can infer the binary fraction that best matches the observed data. 
This approach allows us to estimate the overall binary fraction without resolving individual binaries.

In the following sections, we provide a detailed explanation of this method, including how we simulate binary star populations and generate expected LOSV variability distributions.
We also present the likelihood function used to compare the simulated and observed data, as well as any modifications we have made to the original method to represent the actual binary population in UMi better.

\subsection{Binary Orbital Parameters}
\label{subsec:parameters}

The LOSV of the primary star relative to the center of mass in a binary system is given by the following equation:
\begin{equation}
\label{equ:losv}
    v_{\mathrm{los}}=\frac{q\ \mathrm{sin}\ i}{\sqrt{1-e^2}}\left(\frac{2\pi G m_1}{P(1+q)^2}\right)^{1/3}(\mathrm{cos}\ (\theta+\omega)+e\ \mathrm{cos}\ \omega),
\end{equation}
where $q=m_2/m_1$ is the mass ratio; $m_1$ is the mass of the primary while $m_2$ is the mass of the companion; $i$ is the angle of inclination; $e$ is the orbital eccentricity; $P$ is the period of the binary system; $\theta$ is the true anomaly; and $\omega$ is the argument of periastron. They are the orbital parameters for a given binary system. 

We adopt Equation~\ref{equ:losv} to simulate the LOSV at different epochs of a given star (see more details in Section~\ref{ssec:MCS} below). 
To achieve this goal, we have to know the probability distribution of the seven orbital parameters, i.e., $m_1$, $q$, $P$, $e$, $i$, $\omega$, and $\theta$.

Among these parameters, $q$, $m_1$, $e$, and $P$ are intrinsic, while the other three angles of $i$, $\theta$, and $\omega$ depend on the observer's perspective.
We adopt the distributions of the intrinsic parameters ($q$, $e$, $P$) from two different models developed by \citet{1991A&A...248..485D}(hereafter \citetalias{1991A&A...248..485D}) and \citet{2017ApJS..230...15M}(hereafter \citetalias{2017ApJS..230...15M}), which constrain the binary orbital parameter distributions based on solar neighborhood observations.
The \citetalias{1991A&A...248..485D} model is widely used in many previous studies.
It provides independent distributions for each orbital parameter and is based on solar-like stars, making it straightforward to apply. 
It is important to note that our member stars in UMi are all giant stars with metallicity of $\mathrm{[Fe/H]}<-1$, significantly lower than those in the solar neighborhood.
This means that by adopting the \citetalias{1991A&A...248..485D} model, we are assuming that the binary properties of giants in dwarf galaxies are similar to those of solar-like stars in the solar neighborhood. 

The \citetalias{2017ApJS..230...15M} model considers the joint distributions of various orbital parameters based on more recent observations. Unlike the \citetalias{1991A&A...248..485D} model, which assumes an independent distribution, a few recent studies have reported that different orbital parameters are correlated with each other \citep[e.g.][]{2023A&A...674A..34G}. 
Incorporating these correlations through joint distributions is necessary to achieve a more realistic and comprehensive simulation of binary systems.
In addition, the \citetalias{2017ApJS..230...15M} model also covers a wider range of stellar mass or stellar type, allowing the model to represent a more diverse binary population.
However, in our study, when using the \citetalias{2017ApJS..230...15M} model, we only need to focus on solar-mass stars because the mass of giant stars is close to 1$\msun$. 
Note that the \citetalias{2017ApJS..230...15M} model is also based on solar neighborhood observations. 
Thus, by adopting either the \citetalias{1991A&A...248..485D} model or the \citetalias{2017ApJS..230...15M} model, we are assuming that the binary orbital parameter distributions are the same as in the solar neighborhood. 
This is so far the best we can do.

We refer the readers to the original \citetalias{1991A&A...248..485D} and \citetalias{2017ApJS..230...15M} papers for more details, and in the following, we briefly describe the model distribution adopted for different binary orbital parameters based on the two models.

\subsubsection{Mass of Primary, $m_1$}

For the mass of primary, $m_1$, \citetalias{2018AJ....156..257S} assigned a fixed value of $m_1 = 0.8 M_\odot$ because most stars in UMi are extremely old and are supposed to be located along the red-giant branch (RGB), where the masses are distributed in a very narrow range.
However, we aim to adopt a more realistic mass distribution. 
To achieve this, we sample masses from an isochrone given by the PARSEC tracks\footnote{\url{http://stev.oapd.inaf.it/cgi-bin/cmd}} \citep{2012MNRAS.427..127B}.
We select PARSEC version 1.2S \citep{2014MNRAS.444.2525C,2015MNRAS.452.1068C,2018MNRAS.476..496F} rather than the latest 2.0 to obtain an isochrone with a lower metallicity of [M/H]$=-2.2$, which is the lower limit of the PARSEC tracks but corresponds to the median metallicity of our sample.
The age is set to 10 Gyr, reflecting the average age of UMi \citep{2002AJ....123.3199C}. 
We clip the faint end of the isochrone at an absolute magnitude limit of $M_G>1.46$ mag, which is derived from the faintest star in our data with an apparent magnitude of $m_G=20.92$ and the distance to UMi.
Effectively, the magnitude limit already excludes the fainter, compacter parts such as main-sequence stars and dwarfs, which suggests a surface gravity limit of $\log g<3.0$
Additionally, we also remove the horizontal branch, as we have done with the real data.

We randomly assign the initial mass of each mock star using the initial mass function derived by \citet{2002Sci...295...82K}.
Given the initial mass, we then extract the corresponding current mass, surface gravity, and G-band magnitude from the isochrone.
To ensure consistency with the observational data, we apply a magnitude selection function to enforce the same G-band magnitude distribution in the mock sample as observed in our real dataset.
The mass is distributed around $0.8 M_\odot$, with outliers contributing only a small fraction. 
In contrast, surface gravity spans a much wider range from 0.9 to over 3.0, which differs significantly from the assumption made by \citetalias{2018AJ....156..257S} of a constant value of $1.0$.
This broader range is mainly due to the depth of DESI observation, which includes fainter/compacter objects. 
Since surface gravity plays an important role when generating the orbital period, it is necessary to adopt the more reliable values obtained from the isochrone.

\subsubsection{Mass Ratio, q}

The mass ratio is defined as $q=m_2/m_1$, where $m_1$ and $m_2$ are the masses of the primary star and the companion.
Here we have the assumption that the companion is an invisible star.
$q$ is limited to be in the range of $0.1<q<1$ in our analysis.

\citetalias{1991A&A...248..485D} takes a normal distribution of $q$ as
\begin{equation}
    \frac{dN}{dq}\propto \mathrm{exp}\left(-\frac{(q-\mu_q)^2}{2\sigma_q^2}\right),
\end{equation}
where $\mu_q=0.23$ and $\sigma_q=0.42$.

\citetalias{2017ApJS..230...15M} model the distribution as
\begin{equation}
    \frac{dN}{d q}\propto q^\gamma,
\end{equation} 
where
\begin{equation}
    \gamma=\begin{cases}
         \gamma_\mathrm{small}, &  0.1<q<0.3; \\
         \gamma_\mathrm{large}, &  0.3<q<1.0,
    \end{cases}
\end{equation}
where both $\gamma_\mathrm{small}$ and $\gamma_\mathrm{large}$ depend on the mass of primary, $m_1$, and the orbital period, $P$.

In this work, we consider the case where $m_1<1.2M_\odot$ and $0<\log P<8.0$. Therefore, $\gamma_\mathrm{small}$ and $\gamma_\mathrm{large}$ can be determined as 
\begin{equation}
\begin{aligned}
\gamma_\mathrm{small}&=0.3 \\
\gamma_\mathrm{large}&=\begin{cases}
         -0.5, &  0<\log P<5.0; \\
         -0.5-0.3(\log P-5.0), &  5.0<\log P<8.0.
         \end{cases} 
\end{aligned}
\end{equation}

Here we also consider the excess possibility of twin binaries, which has been reported to have a clear narrow peak in the probability distribution of the mass ratio at $q>0.95$ for close binaries \citep{2000A&A...360..997T,2006ApJ...639L..67P,2010ApJS..190....1R}. The excess probability is defined as
\begin{equation}
\mathcal{F}_{\mathrm{twin}}=\begin{cases}
     \mathcal{F}_{\mathrm{twin,1}},&  0<\log P<1.0; \\
     \mathcal{F}_{\mathrm{twin,1}}\left(1-\frac{\log P-1}{\log P_{\mathrm{twin}}-1}\right),&  1.0<\log P<\log  P_{\mathrm{twin}};\\
     0, &  \log P_{\mathrm{twin}}<\log P,
\end{cases}
\end{equation}
where 
\begin{equation}
\begin{aligned}
\log P_{\mathrm{twin}}&=8.0-\frac{m_1}{M_\odot}\\
\mathcal{F}_{\mathrm{twin,1}}&=0.3-0.15\log\left(\frac{m_1}{M_\odot}\right).
\end{aligned}
\end{equation}

We present only the relevant parts of the distribution applicable to our study. The reader can refer to \citetalias{2017ApJS..230...15M} for detailed information on the complete formulations for other primary mass ranges.

\subsubsection{Period, P}
\label{sssec:P}

The adopted orbital period distribution most significantly affects the binary fraction constraints in our analysis \citep[][\citetalias{2018AJ....156..257S}]{2013ApJ...779..116M,2017AJ....153..254S}.
\citetalias{1991A&A...248..485D} model the orbital period distribution with a log-normal form
\begin{equation}
\frac{dN}{d\log P}\propto \mathrm{exp}\left(-\frac{(\log P-\mu_{\log P})^2}{2\sigma_{\log P}^2}\right).
\end{equation}
 
We adopt $\mu_{\log P}=4.8$ and $\sigma_{\log P}=2.3$, where the period is in units of day. The adopted mean and standard deviation values provide the best fit to F7 to G9 type stars in our solar neighborhood, as found by \citetalias{1991A&A...248..485D}. 
We also investigated different values from several other studies, including \citet{2010ApJS..190....1R}, \citet{1992ApJ...396..178F}, and \citet{2011MNRAS.417.1702M}.
A larger mean value of the period decreases the inferred binary fraction, whereas a larger standard deviation increases it. 
Specifically, a larger standard deviation leads to a higher fraction of short-period binaries, which have a significant effect on the LOSV variabilities (see more details in Section~\ref{ssec:beta} below). 
In contrast, medium- to long-period binaries have minimal impact on the results due to the fact that our UMi member star observations are at most spanned by one year, which cannot provide strong constraints on medium- or long-period binaries. 
In this work, we focus on the values from \citetalias{1991A&A...248..485D} which have been demonstrated as the best model among the works considered above, as supported by \citetalias{2018AJ....156..257S}.

Using Kepler's third law, we can constrain the minimum and maximum periods based on the semi-major axis, $a$.
We adopt the same expression for the maximum semi-major axis as used by \citetalias{2018AJ....156..257S}, which is considered as the mean free path for stellar collision,
\begin{equation}
    a_{\mathrm{max}}=(\pi\sigma_v t \lambda)^{-1/2}\simeq1.26\times10^{14} \mathrm{m},
\end{equation}
where $\sigma_v=8.0$ km$^{-1}$ is the velocity dispersion of UMi (\citetalias{2018AJ....156..257S}), $t=10$ Gyr is the average age of its member stars \citep{2002AJ....123.3199C}, and $\lambda=0.2344$ pc$^{-3}$ is the stellar number density, 
which is derived by the central luminosity density $L=0.006L_\odot$ pc$^{-3}$ \citep{1998ARAA..36..435M} and the assumption that the average mass of the stars in UMi is $0.4M_\odot$, applying the mass-luminosity relation $L/L_\odot\propto (M/M_\odot)^4$.
The corresponding maximum limit of the period is around $10^7$ days.

The lower limit of the semi-major axis is considered as the radius of the Roche lobe of the primary, which depends on the mass ratio $q$.
The analytical formula of the approximate Roche radius is given by \citet{1983ApJ...268..368E},
\begin{equation}
    \frac{r_L}{A}=\frac{0.49q_r^{2/3}}{0.6q_r^{2/3}+\mathrm{ln}(1+q_r^{1/3})},
\end{equation}
where $q_r=1/q$ is the inverse of the mass ratio in this work, $r_L$ is the radius of the sphere whose volume approximates the Roche lobe of the primary, and $A=a\times(1-e)$ is the orbital separation of this binary system, where $e$ is the eccentricity given in the following subsection.
Besides, we also require the radius $R=\sqrt{\mathrm{G}m_1/g}$ of the primary to be smaller than the Roche lobe, i,e, $r_L > R$; otherwise, the star is considered to be a single star.
Therefore, the lower limit of the semi-major axis is determined by the combined constraints $r_L > R$ and $a>R$.
These conditions are imposed through our sampling process rather than derived analytically. 
As a result, the corresponding orbital period has a lower limit of approximately 4 days.

On the other hand, \citetalias{2017ApJS..230...15M} demonstrates a more complicated distribution for the orbital period as a function of the primary mass $m_1$. We do not repeatedly show the form here, and the readers can check \cite{2017ApJS..230...15M} for details. 
These values are all based on their measurements, combining numerous observations. 
The equations presented above are valid for systems with mass ratios of $q>0.3$. 
For systems with smaller mass ratios $0.1<q<0.3$, the overall distribution remains qualitatively similar, exhibiting only minor amplitude variations that do not impact our sampling.
Therefore, to simplify the simulation, we adopt the same analytical functions to describe systems across the entire range of mass ratios $0.1<q<1.0$, as the variations are not significant enough to be distinct.

\subsubsection{Eccentricity, e}

\citetalias{2018AJ....156..257S} choose a uniform distribution to simplify the model since the eccentricity only has a small effect on the observed binary fraction.
\begin{equation}
    \frac{dN}{de}\propto \mathrm{const}.
\end{equation}
The eccentricity can range from 0 to $e_{\mathrm{max,1}}=1-(R/a)$, where $a$ is the semi-major axis and $R$ is the radius of the primary mentioned above. This limit is set to prevent the stars from potential collisions.

\citetalias{2017ApJS..230...15M} investigated a power law distribution for the eccentricity,
\begin{equation}
    \frac{dN}{de}\propto e^\eta,
\end{equation}
where $\eta$ is a parameter depending on both $m_1$ and $P$.
But we only have a single form based on $P$ due to our narrow mass range,
\begin{equation}
\eta=
\begin{cases}
    0,&\log P<0.5;\\
    0.6-\frac{0.7}{\log P-0.5},&0.5<\log P<6.0;\\
    0.6-\frac{0.7}{6.0-0.5},&6.0<\log P.
\end{cases}
\end{equation}
They set another upper limit for the eccentricity as $e_{\mathrm{max,2}}=1-(P/2)^{-2/3}$, 
which guarantees the Roche lobe filling factor to be approximately less than $70\%$. We pick up the smaller one between this value and the one mentioned above to be the final upper limit, $e_{\mathrm{max}}=\mathrm{min}(e_{\mathrm{max,1}},e_{\mathrm{max,2}})$.

\subsubsection{Inclination, i}

The remaining three parameters are not model-dependent.
Instead, they are determined by the orientation of the binary system relative to the observer and are assigned based on the actual observational geometry.

The inclination angle $i$ is proportional to $\mathrm{sin}\ i$ while assuming a random orientation,
\begin{equation}
    \frac{dN}{di}\propto \mathrm{sin}\ i.
\end{equation}
It ranges from $0$ (face-on) to $\pi/2$ (edge-on).

\subsubsection{Argument of Periastron, $\omega$}

The argument of periastron $\omega$ is the angle between the ascending node of the orbit and the periastron point.
It follows a uniform distribution from 0 to $2\pi$,
\begin{equation}
    \frac{dN}{d\omega}\propto \mathrm{const}.
\end{equation}

\subsubsection{True Anomaly, $\theta$}

The true anomaly is an angular parameter that describes the position of an object moving along its orbital path. 
Specifically, it is the angle between the direction of the periastron and the current position of the object, measured from the focus of the orbit.
The true anomaly depends on the orbital period and eccentricity; however, it lacks an analytical solution. 
Therefore, numerical methods are typically employed to calculate the true anomaly. 

Firstly, we define the mean anomaly, $M$, which follows a uniform distribution from 0 to $2\pi$,
\begin{equation}
    \frac{dN}{dM}\propto \mathrm{const}.
\end{equation}

Next, we numerically solve the eccentric anomaly $E$, which is also an angular parameter, with the following form,
\begin{equation}
    M=E-e\ \mathrm{sin}\ E.
\end{equation}

Finally, we obtain the true anomaly $\theta$ using the following relation,
\begin{equation}
    \theta=2\mathrm{arctan}\left(\sqrt{\frac{1+e}{1-e}}\mathrm{tan}\frac{E}{2}\right).
\end{equation}

For each object, we have a series of observations with time separations. 
We randomly assign an initial mean anomaly $M_0$ for the first observation.
The mean anomalies for subsequent observations are then calculated using the corresponding time separation $\Delta t$ and the orbital period $P$, i.e., $M_i=M_0+2\pi\Delta t/P$.

\subsection{Monte Carlo Simulations}
\label{ssec:MCS}

Based on the distribution of all seven orbital parameters, we can perform Monte Carlo simulations to generate realizations of observations at different epochs and for each star, corresponding to our real member star sample of UMi, treating it as a binary or non-binary sample:
\begin{enumerate}
    \item \textbf{Generate Orbital Parameters:}
     For a star in an assigned binary sample, we randomly generate the seven orbital parameters based on the distributions mentioned above. The six orbital parameters, excluding the true anomaly, remain the same for each star across all observations. The true anomaly varies according to the actual observation epochs.
     If the star is considered in a non-binary sample, we skip this step.
     
    \item \textbf{Calculate LOSV:}
     Starting from a randomly assigned initial orbital phase, we calculate the LOSV at subsequent epochs using the orbital parameters and Equation~\ref{equ:losv}. Each mock star is associated with a real observed star in our sample, and its simulated LOSV measurements are generated at the same observational epochs as the corresponding real star. The temporal sampling of the mock data accurately reflects the true time baselines of the observations, allowing for a realistic comparison between simulated and observed LOSV variability.
     For non-binary stars, the LOSV is set to 0 $km s^{-1}$. Since our focus is on LOSV variability rather than absolute velocities, we adopt a systemic LOSV of 0 $km s^{-1}$ for both binary and non-binary stars for the mock data.
     
    \item \textbf{Add Gaussian Noise:}
    We add Gaussian random noise to the LOSV of both binary and non-binary stars (with the LOSV centered at zero for non-binary stars), using the quadrature sum of the measurement uncertainty, $\sigma_{\rm{obs}}$, and the fixed systematic error floor, $\sigma_{\rm{sys}}$, as the standard deviation, $\sigma_{v_{\rm{los}}}=\sqrt{\sigma_{{\rm{obs}}}^2+\sigma_{\rm{sys}}^2}$. When $\sigma_{\rm{sys}}$ is not yet fixed in the analysis, it is treated as a variable parameter and adjusted accordingly during the simulation process (see Section~\ref{ssec:syserr} for further details).
    
    \item \textbf{Repeat Simulations:}
    We repeat the above procedure $n = 10^5$ times to generate UMi-like mock samples, each consisting of 670 stars with 2,147 LOSV measurements, for both the binary and non-binary cases. Finally, we obtain $10^5$ simulated datasets in which all stars are assumed to be binaries and another $10^5$ dataset in which all stars are assumed to be non-binaries. Each simulation represents a realization based on the actual member star sample of UMi.

\end{enumerate}
Based on the above steps, we generate numerous mock catalogs with simulated LOSVs for each star and at its different observational epochs, as well as including their corresponding observational uncertainties. 
As mentioned earlier, we treat the stars in the dwarf galaxy as a collective ensemble rather than attempting to identify individual binary systems. 
Therefore, we compare the LOSV variability distributions of each simulated sample with the observational data to determine the binary fraction that best fits the observations. We introduce the definition of the LOSV variability in the next subsection.

\subsection{LOSV Variability, $\beta$}
\label{ssec:beta}

Following \cite{2018AJ....156..257S}, we can define the LOSV variation for each pair of observations as
\begin{equation}
    \alpha_{i,j}=\frac{|v_{\mathrm{los},i}-v_{\mathrm{los},j}|}{\sqrt{\sigma_{v_\mathrm{los},i}^2+\sigma_{v_\mathrm{los},j}^2}},
\label{equ:alpha}
\end{equation}
where $v_\mathrm{los}$ is the LOSV and $\sigma_{v_\mathrm{los}}$ is the corresponding uncertainty. 
The subscripts $i$ and $j$ denote different observations of the same star. Thus, for a star with multiple observations, we calculate an $\alpha$ value for each unique pair of observations. Consequently, a star with $n$ observations contributes $n(n-1)/2$ pairs with different time separations to the $\alpha$ distribution. 
Notably, before constraining the systematic error floor in our analysis, the uncertainties in the denominator include only the observational errors. Once the systematic error floor is determined, we repeat the analysis by incorporating it into the total uncertainty in quadrature. Further details on this procedure can be found in Section~\ref{ssec:syserr}.

However, this method would artificially inflate the influence of stars with more observations. Stars with a higher number of epochs contribute more $\alpha$ values that are not independent, potentially underestimating the uncertainty. To mitigate this, we further adopt a modified approach to ensure that each star contributes equally to the final distribution of the LOSV variability. Specifically, for each star, we define a single representative LOSV variability as
\begin{equation}
    \beta=\sqrt{\frac{2}{n(n-1)}\sum_{i<j}\alpha_{i,j}^2},
\label{equ:beta}
\end{equation}
and use this quantity in the subsequent Bayesian analysis (see Section~\ref{sec:likelihood}).

The distribution of $\beta$ represents the overall LOSV variability within the entire sample. For a dwarf galaxy system, the time scale of stellar motions in the potential well of the dwarf galaxy itself is much longer than the time scale associated with binary orbital motions. Thus, the observed LOSV variability, quantified by the distribution of $\beta$, is dominated by the binary orbital motion.
In a dwarf galaxy with a higher binary fraction, we expect the $\beta$ distribution to exhibit larger values, reflecting the increased LOSV variability due to binary motion. 
Conversely, in galaxies with a low binary fraction, fewer stars exhibit significant LOSV variability over the observed time baselines, leading to a $\beta$ distribution that peaks at smaller values.

\subsection{Likelihood Function}
\label{sec:likelihood}

\begin{figure}
\plotone{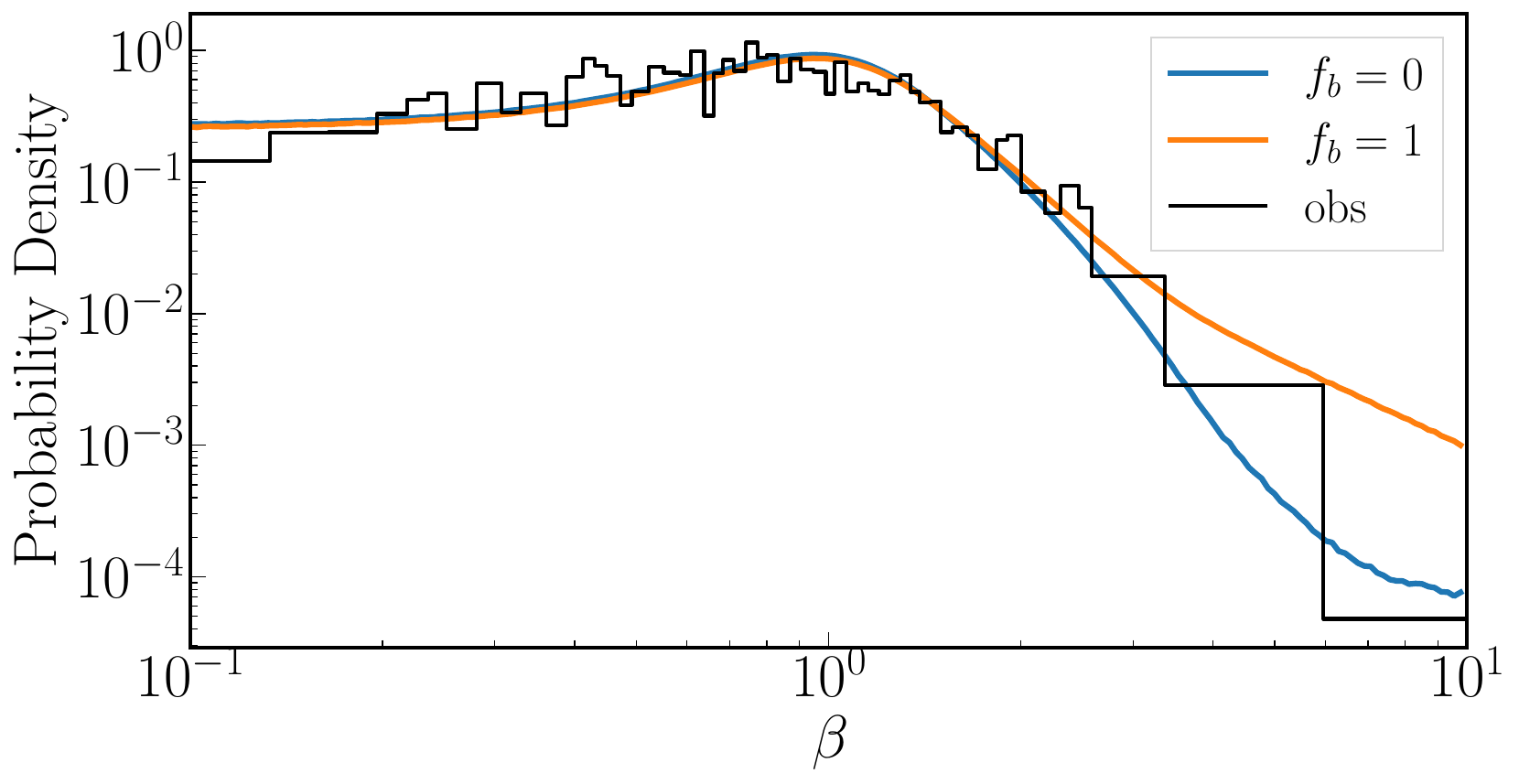}
\caption{The distribution of LOSV variability, $\beta$, as constructed from our $10^5$ mock samples. The blue line represents the probability density profile derived from the non-binary samples, while the orange line corresponds to that from the binary samples. The black histogram shows the observed $\beta$ distribution of the real data.
\label{fig:beta_dist}}
\end{figure}

As described in Section~\ref{ssec:MCS}, we derive the $\beta$ distributions for both the binary and non-binary mock star samples, as shown in Figure~\ref{fig:beta_dist}. The probability distributions of $\beta$ are constructed from all $10^5$ mock realizations and normalized to unity. The black histogram represents the $\beta$ distribution obtained from the observational data. By combining the binary and non-binary samples with varying fractions, we can infer the binary fraction that best reproduces the observed distribution.

To be more specific, given a binary fraction $f_b$, we estimate the probability of observing the LOSV variability, $\beta_{i, \rm obs}$, for each star in our sample as follows, 
\begin{equation}
\begin{aligned}
    P(\beta_{i, \rm obs}|f_b)=&f_b\times P(\beta_{i, \rm obs}|f_b=1)\\
    &+(1-f_b)\times P(\beta_{i, \rm obs}|f_b=0),
\end{aligned}
\end{equation}
where $P(\beta_{i, \rm obs}|f_b=1)$ and $P(\beta_{i, \rm obs}|f_b=0)$ represent the probability of $\beta_{\rm obs}$ in the binary and non-binary populations, respectively, as derived from our mock simulations.

Then the overall likelihood for a given binary fraction, $f_b$, is equal to the product of these probabilities from each star,
\begin{equation}
    \mathcal{L}(f_b|\beta_{\rm obs}) = \prod_i P(f_b,\beta_{i, \rm obs}) .
\end{equation}

We apply a Markov Chain Monte Carlo (MCMC) approach to estimate the binary fraction and its associated uncertainty. For this purpose, we use the \textsc{python} package \textsc{emcee}\footnote{\url{https://github.com/dfm/emcee}} \citep{emcee}, an affine-invariant ensemble sampler designed for Bayesian parameter estimation. In this analysis, we construct the posterior probability distribution by combining the likelihood function derived from the observed $\beta$ distribution with a uniform prior to the binary fraction. The final posterior distribution provides both the best-fit binary fraction and its credible interval, which we report as the 68\% highest posterior density interval (HPDI).

\subsection{Determination of Systematic Error Floor of LOSV}
\label{ssec:syserr}

It is necessary to determine the appropriate value of the systematic error floor in LOSV for our sample. Although treating the systematic error floor as another free parameter, in addition to the binary fraction, within the MCMC framework would be ideal in principle, such an approach is computationally expensive and therefore impractical for our current analysis.

To overcome this limitation, we adopt a grid-based strategy. We vary the systematic error floor $\sigma_{\rm sys}$ from 0.8 to 2.0 km~s$^{-1}$ in increments of 0.1 km~s$^{-1}$. For each value, we recalculate the corresponding $\beta$ distribution by adding Gaussian noise to the simulated LOSVs, using the total uncertainty, $\sigma_{v_{\rm{los}}}=\sqrt{\sigma_{{\rm{obs}}}^2+\sigma_{\rm{sys}}^2}$ as described in Step 3 of Section~\ref{ssec:MCS}. We then perform the MCMC estimation of the binary fraction based on each recalculated $\beta$ distribution.

\begin{figure}
\plotone{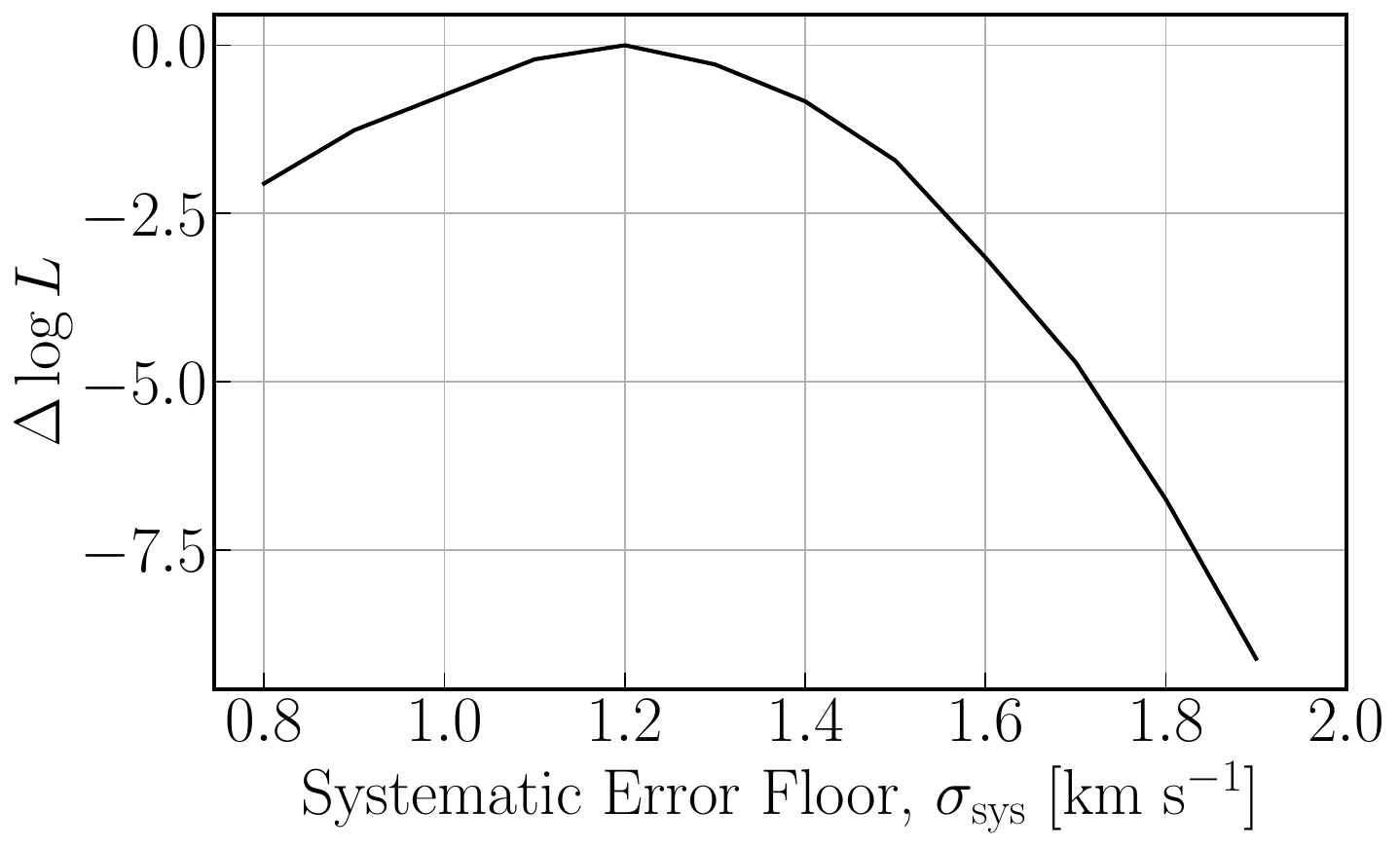}
\caption{Relative maximum log-likelihood as a function of the assumed systematic error floor ($\sigma_{\rm sys}$). Each point represents the maximum log-likelihood obtained from the MCMC analysis for a given value of $\sigma_{\rm sys}$, normalized by subtracting the overall maximum log-likelihood. The peak occurs at $\sigma_{\rm sys} = 1.2\ {\rm km\ s}^{-1}$, indicating the most likely value of the systematic error floor under the observing conditions of our sample.
\label{fig:syserr}}
\end{figure}

To identify the most suitable value of the systematic error floor, we compare the maximum log-likelihood values obtained from the MCMC samplers for each $\sigma_{\rm sys}$. 
Figure~\ref{fig:syserr} presents the relative maximum log-likelihoods for different systematic error floor values, with the values normalized by subtracting the absolute maximum log-likelihood to highlight the relative differences. 
As shown in the figure, the maximum log-likelihood is achieved when the systematic error floor is $\sigma_{\rm sys} = 1.2\ {\rm km\ s}^{-1}$, indicating that this value best characterizes the systematic error floor under the observing conditions of our sample.
This value is also consistent with the values estimated in \citet{2025arXiv250514787K}. 
Accordingly, we adopt $\sigma_{\rm{sys}}=1.2\ \rm{km\ s}^{-1}$ as the fixed systematic error floor in our final analysis and use it to construct the posterior distribution for the binary fraction. For all of the analyses in the following sections, the binary fraction is the only free model parameter.

\section{Results}
\label{sec:results}

\subsection{Binary Fraction in UMi}
\label{ssec:bf}
\begin{figure}
\plotone{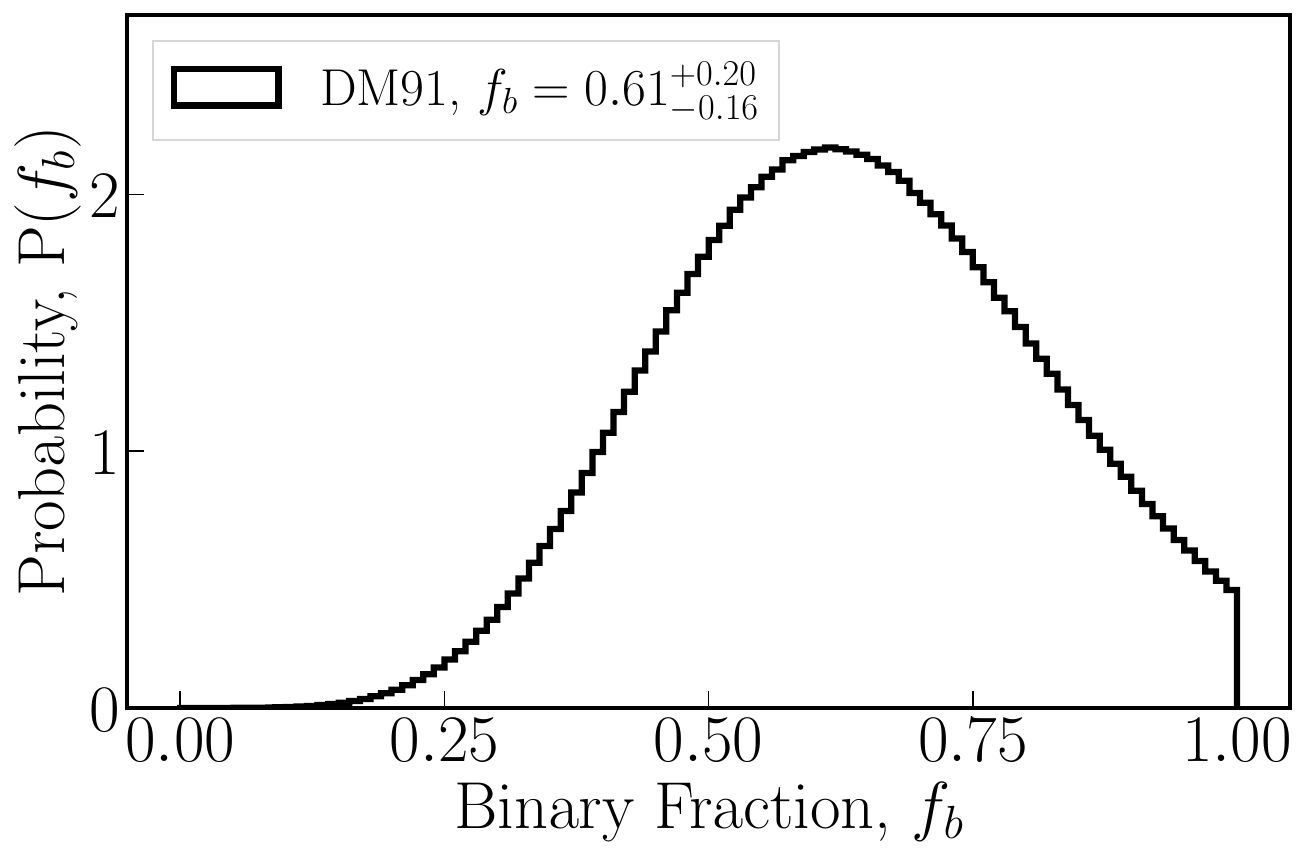}
\plotone{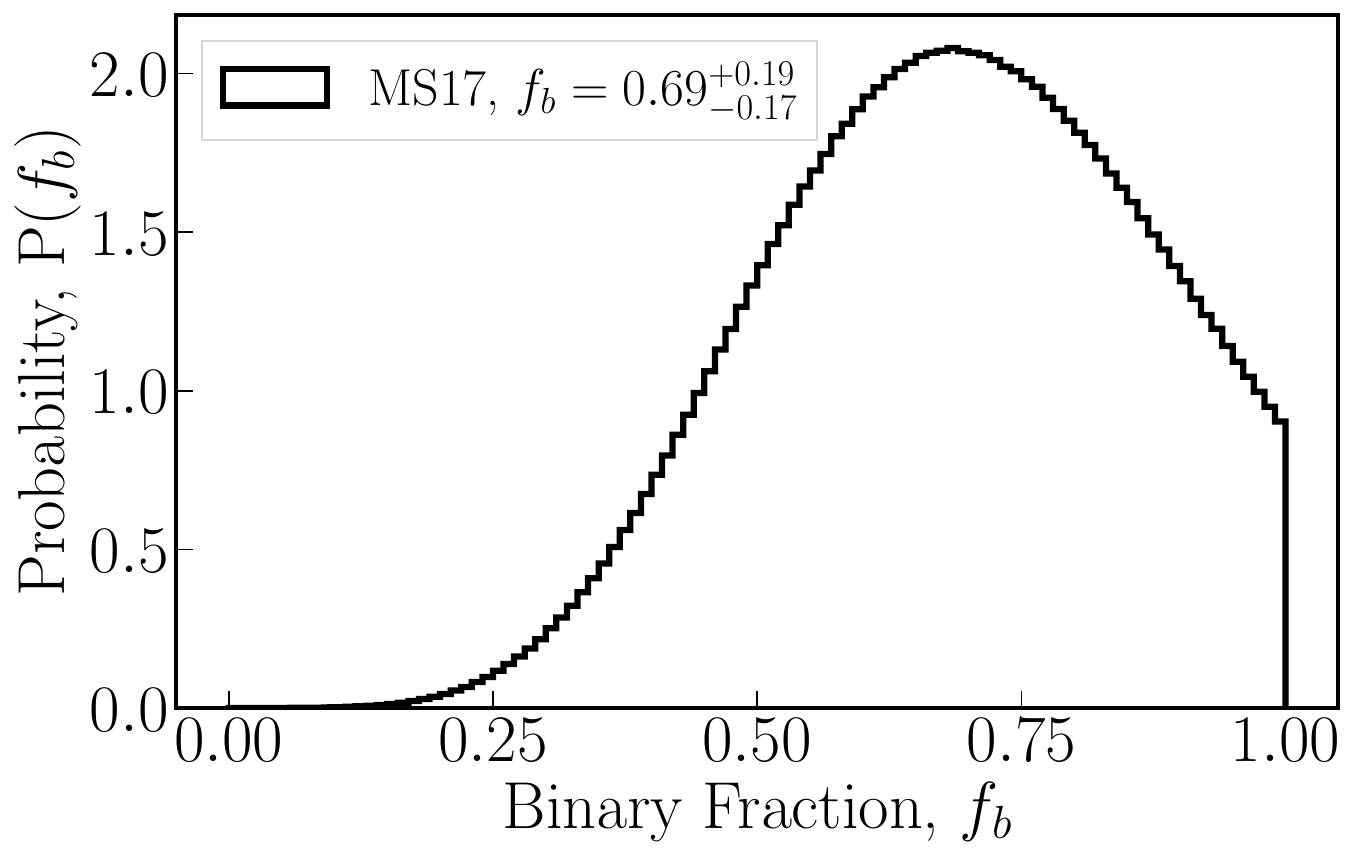}
\caption{The top and bottom panels show the best constraint PPDs of binary fractions based on the \citetalias{1991A&A...248..485D} and \citetalias{2017ApJS..230...15M} models, respectively. 
The blue dashed line indicates the median and the red dashed lines show the $1\sigma$ range.
\label{fig:results}}
\end{figure}

We derive the binary fraction of UMi using both \citetalias{1991A&A...248..485D} and \citetalias{2017ApJS..230...15M} models.
The posterior probability distribution (PPD) is shown in Figure~\ref{fig:results}.
From the \citetalias{1991A&A...248..485D} model, we obtain a binary fraction of $0.61^{+0.16}_{-0.20}$, while the  \citetalias{2017ApJS..230...15M} model yields a result of $0.69^{+0.19}_{-0.17}$.
The quoted $1\sigma$ uncertainties are derived from the 68\% HPDI of the cumulative PPDs.
While \citet{1996AJ....111..750O} reported an early, highly uncertain estimate of the binary frequency near one-year periods in UMi, the expanded multi-epoch LOSV dataset and improved binary population modeling in this work yield substantially more robust constraints on the total binary fraction.
Our results from the \citetalias{2017ApJS..230...15M} model agree well with the previous work done by \citetalias{2018AJ....156..257S}, who found a best-fit binary fraction of $0.78^{+0.09}_{-0.08}$.
Although their study also followed the \citetalias{1991A&A...248..485D} framework, our approach integrates a stellar isochrone in the modeling procedure to generate quantities including the mass of the primary star and surface gravity for each star (see Section~\ref{subsec:parameters} for details), providing a more physically motivated representation of the stellar population in UMi.
Our results exhibit larger uncertainties than those reported by \citetalias{2018AJ....156..257S}, primarily due to differences in the definition of the LOSV variability, $\beta$. As discussed in \ref{ssec:beta}, the approach adopted by \citetalias{2018AJ....156..257S} effectively amplifies the contribution of stars with a larger number of observations. Consequently, multiple $\beta$ values from the same star enter the likelihood function, which increases the apparent statistical weight of these stars and leads to an underestimation of the overall uncertainties. On the other hand, our Equation~\ref{equ:beta} forces the same weight from each individual star.

We further subdivide our sample into metal-rich and metal-poor subsamples based on their metallicities. 
For stars with multiple observations, we obtain their metallicity from the stacked spectra (see Section~\ref{sec:umimem}).
We divide stars into metal-rich and metal-poor populations based on their metallicity using a threshold of $[\mathrm{Fe}/\mathrm{H}]=-2.14$.
This value, $-2.14$, corresponds to the median metallicity of the full sample; thereby our metal-rich ($[\mathrm{Fe}/\mathrm{H}]>-2.14$) and metal-poor ($[\mathrm{Fe}/\mathrm{H}]<-2.14$) subsamples contain equal number of stars. 
Explicitly, we have a metal-rich sample of 335 stars with a total of 1,024 observations at different epochs. 
The metal-poor sample also has 335 stars, with a total of 1,123 observations.

\begin{figure}
\plotone{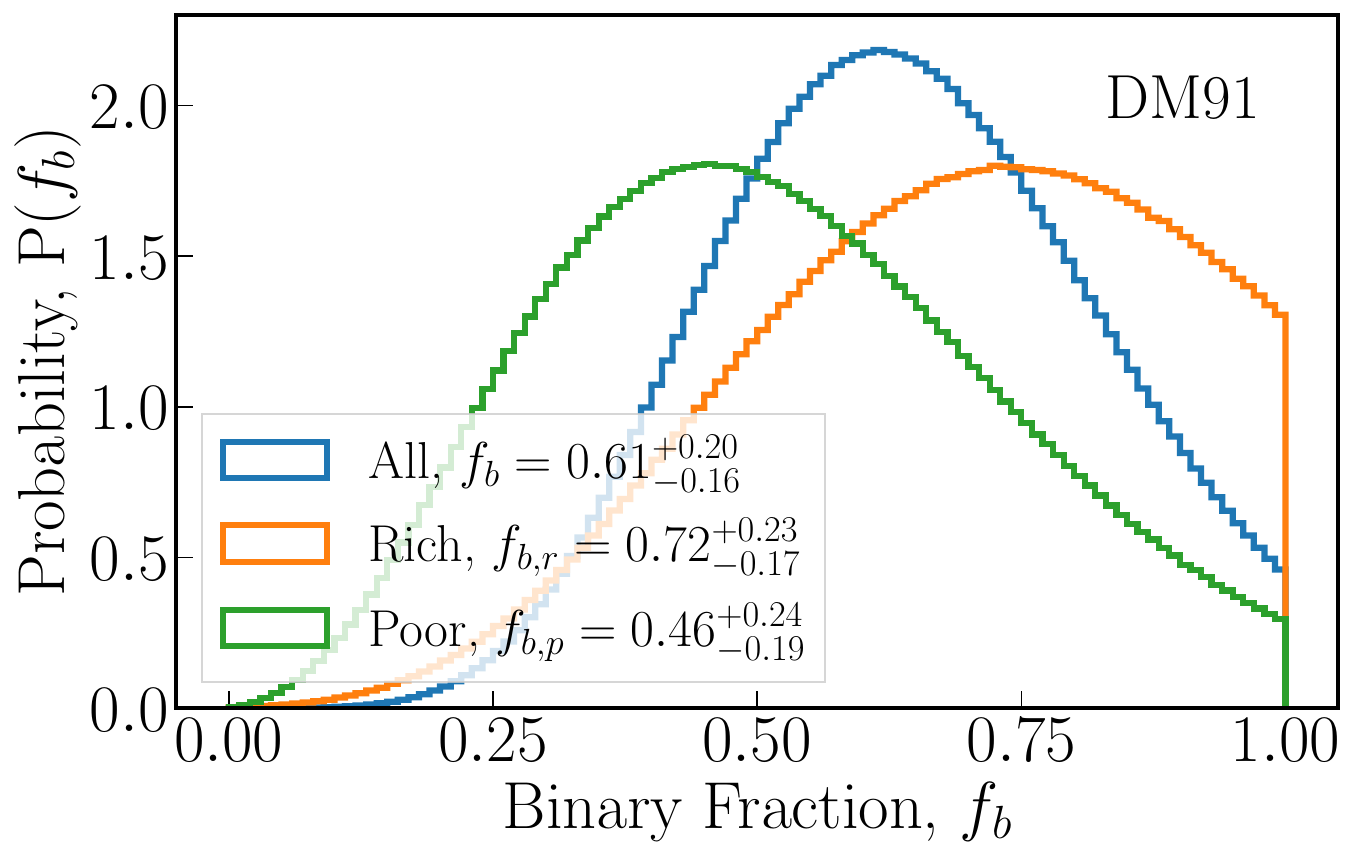}
\plotone{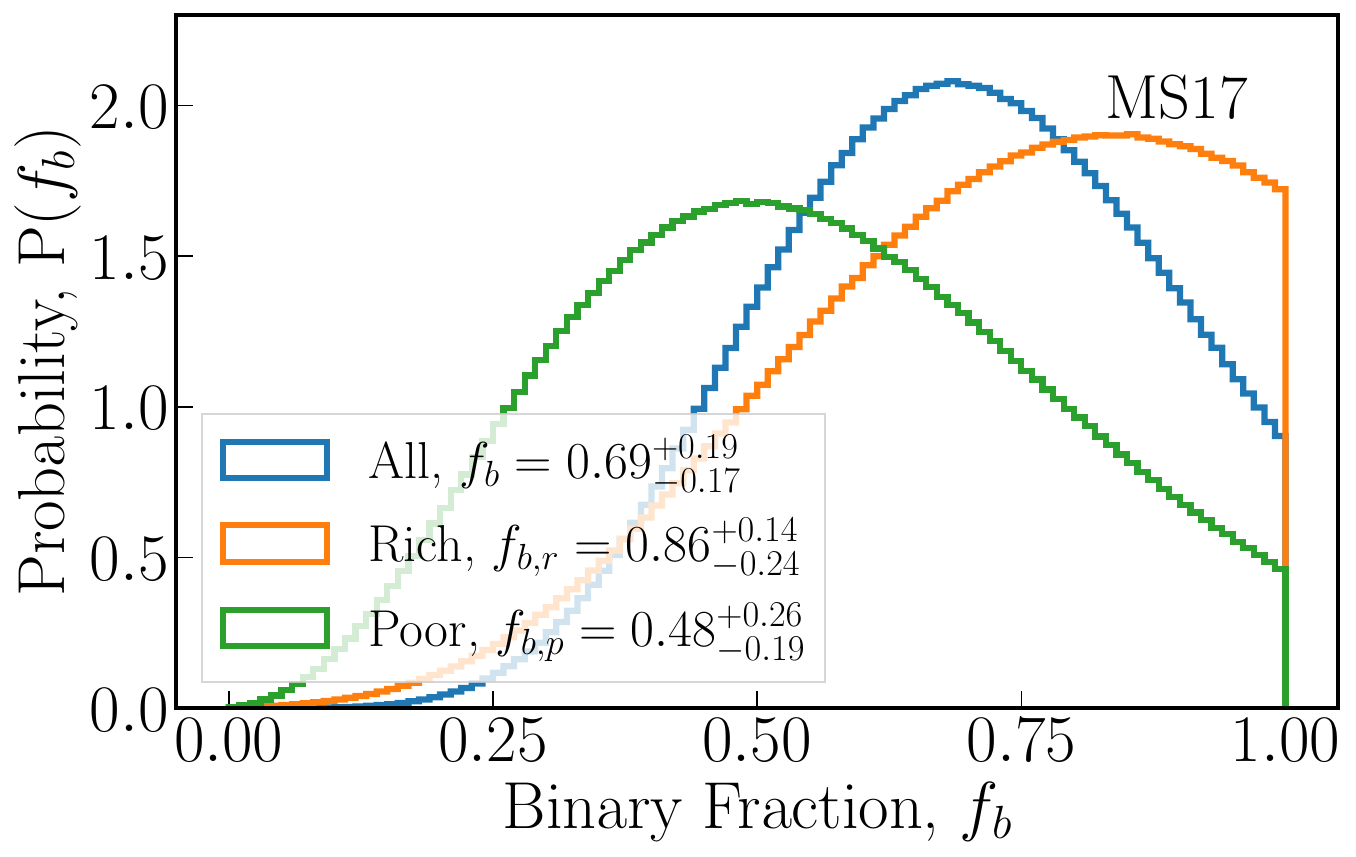}
\caption{We present the PPDs for the metal-rich (orange) and metal-poor (green) populations, as indicated in the legend. The top and bottom panels are based on the \citetalias{1991A&A...248..485D} and \citetalias{2017ApJS..230...15M} models, respectively. The blue curves show the results for the full sample (as in Figure~\ref{fig:results}).
\label{fig:results_comp}}
\end{figure}

We constrain the binary fractions for the metal-rich and metal-poor populations separately, based on the same modeling approach for the full sample. 
As shown in Figure~\ref{fig:results_comp}, these two populations exhibit somewhat different binary fractions. For both the \citetalias{1991A&A...248..485D} and \citetalias{2017ApJS..230...15M} models, the metal-rich sample has a higher binary fraction than the metal-poor sample. 
In the \citetalias{1991A&A...248..485D} model, the metal-rich population has a binary fraction of $0.72^{+0.23}_{-0.17}$, whereas the metal-poor population has a value of $0.46^{+0.24}_{-0.19}$.
Considering the \citetalias{2017ApJS..230...15M} model, the corresponding values are $0.86^{+0.14}_{-0.24}$ and $0.48^{+0.26}_{-0.19}$ for metal-rich and metal-poor populations, respectively. However, the difference between the binary fractions of the metal-rich and metal-poor populations is marginal concerning the uncertainties, which is slightly higher than 1$\sigma$.

\subsection{Selection Effects}
\label{ssec:selecteff}
In Section~\ref{ssec:bf}, we report a marginal indication that the metal-rich stellar population of UMi may exhibit a higher binary fraction than the metal-poor population. In this section, we examine whether this metallicity dependency could arise from sample selection effects rather than reflecting an intrinsic difference. Specifically, we investigate the impact of non-uniform observing strategies across UMi, including variations in the number of epochs and the time baselines or separations between them.
For example, we know that the spatial distributions of metal-rich and metal-poor stars differ, with more metal-rich stars in central regions. The tiling of UMi, however, is spatially varying (see Figure~\ref{fig:footprint}), which may result in different number of epochs and time baselines for different stellar populations and may affect the comparison between different populations.

Moreover, metal-rich stars typically exhibit stronger spectral line features, enabling more precise LOSV measurements, whereas metal-poor stars often yield larger velocity uncertainties. Since LOSV variability (see Equation~\ref{equ:beta}) is a key factor in determining the binary fraction, which can be significantly influenced by LOSV precision, these selection effects have to be controlled and understood first, before concluding that the different binary fractions between metal-rich and metal-poor stars are real.

\begin{figure*}
\plotone{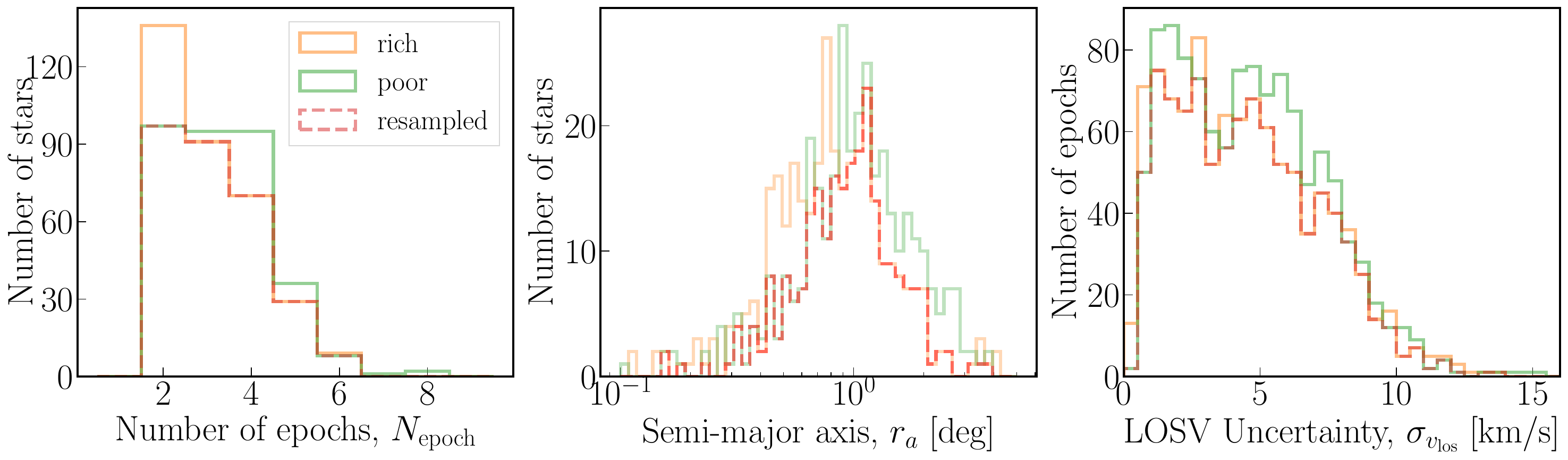}
\plotone{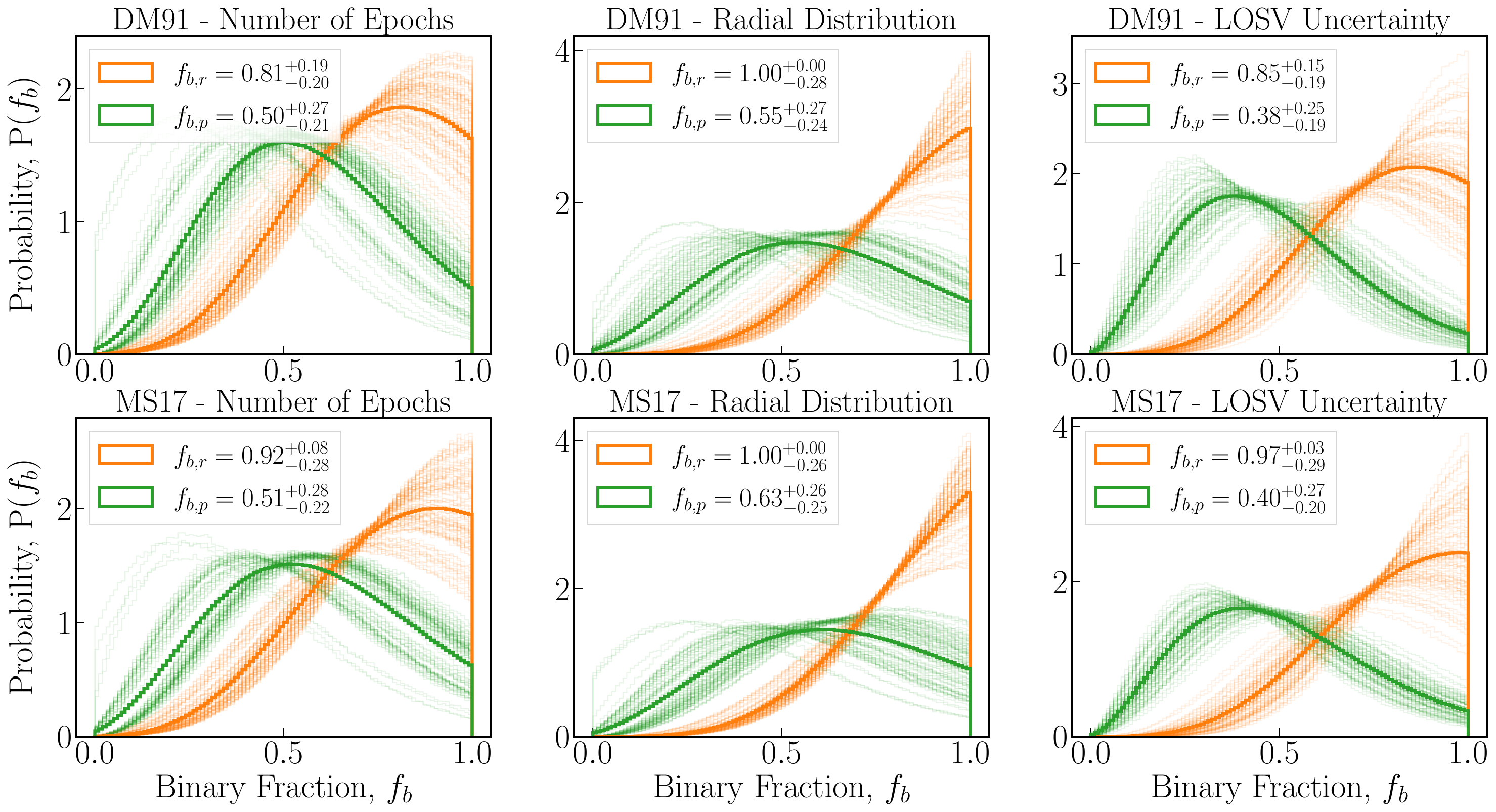}
\caption{Resampling of the metal-rich and metal-poor populations to match their distributions in number of epochs ({\bf left} column), distance to the dwarf center in projection (semi-major axis of elliptical isophotes in number density, {\bf middle} column), and LOSV uncertainty ({\bf right} column), respectively.
{\bf Top:} Original distributions for the metal-rich (orange) and metal-poor (green) samples, along with the matched distribution after resampling (dashed red).
{\bf Middle:} The PPDs of the binary fraction for the \citetalias{1991A&A...248..485D} model following resampling in both populations.
{\bf Bottom:} The corresponding results for the \citetalias{2017ApJS..230...15M} model. 
We perform the resampling procedure 100 times. 
The translucent histograms show the distribution of results from individual realizations, while the bold lines represent the combined PPDs from all trials. 
\label{fig:obs_effect}}
\end{figure*}

As shown in panels of the top row of Figure~\ref{fig:obs_effect}, the metal-rich population in UMi is indeed slightly more centrally concentrated. Stars in the metal-rich subsample (orange histogram) have, on average, slightly fewer epochs of observation than those in the metal-poor subsample (green). 
Moreover, the top right panel of Figure~\ref{fig:obs_effect} shows that the metal-rich population indeed exhibits lower LOSV uncertainties, whereas the LOSV measurements of the metal-poor population are subject to a bit larger uncertainties. 

To mitigate such selection effects, we resample both the metal-rich and metal-poor populations so that they share exactly the same distributions of number of epochs (i.e., the number of epochs per star), projected radius (semi-major axis of elliptical isophotes in number density), and LOSV uncertainty. 

After resampling, the red dashed histograms illustrate that the updated metal-rich and metal-poor subsamples now share identical distributions. To achieve this, we randomly remove stars from either population to ensure they match exactly in key observational properties. Specifically, we perform this resampling procedure 100 times to generate multiple realizations of matched metal-rich and metal-poor samples. In each realization, the two subsamples have identical distributions in terms of number of epochs, radial positions, and LOSV uncertainties, although the specific stars retained differ between realizations.
However, we do not resample all three properties simultaneously, as the remaining number of stars/measurements would be too small to place meaningful constraints on the binary fraction.

The middle (\citetalias{1991A&A...248..485D} model) and bottom (\citetalias{2017ApJS..230...15M} model) rows of Figure~\ref{fig:obs_effect} show the PPDs of the binary fractions for the resampled metal-rich (orange) and metal-poor (green) populations. Note that the solid curve is the overall PPD from the 100 resampled realizations mentioned above, with the results for the 100 realizations shown in the background by curves with thin, lighter colored lines. Although the resampling slightly alters the measured binary fractions and increases the uncertainties due to the reduced sample size, the difference between the binary fractions of the metal-rich and metal-poor populations remains.

\begin{figure}
\plotone{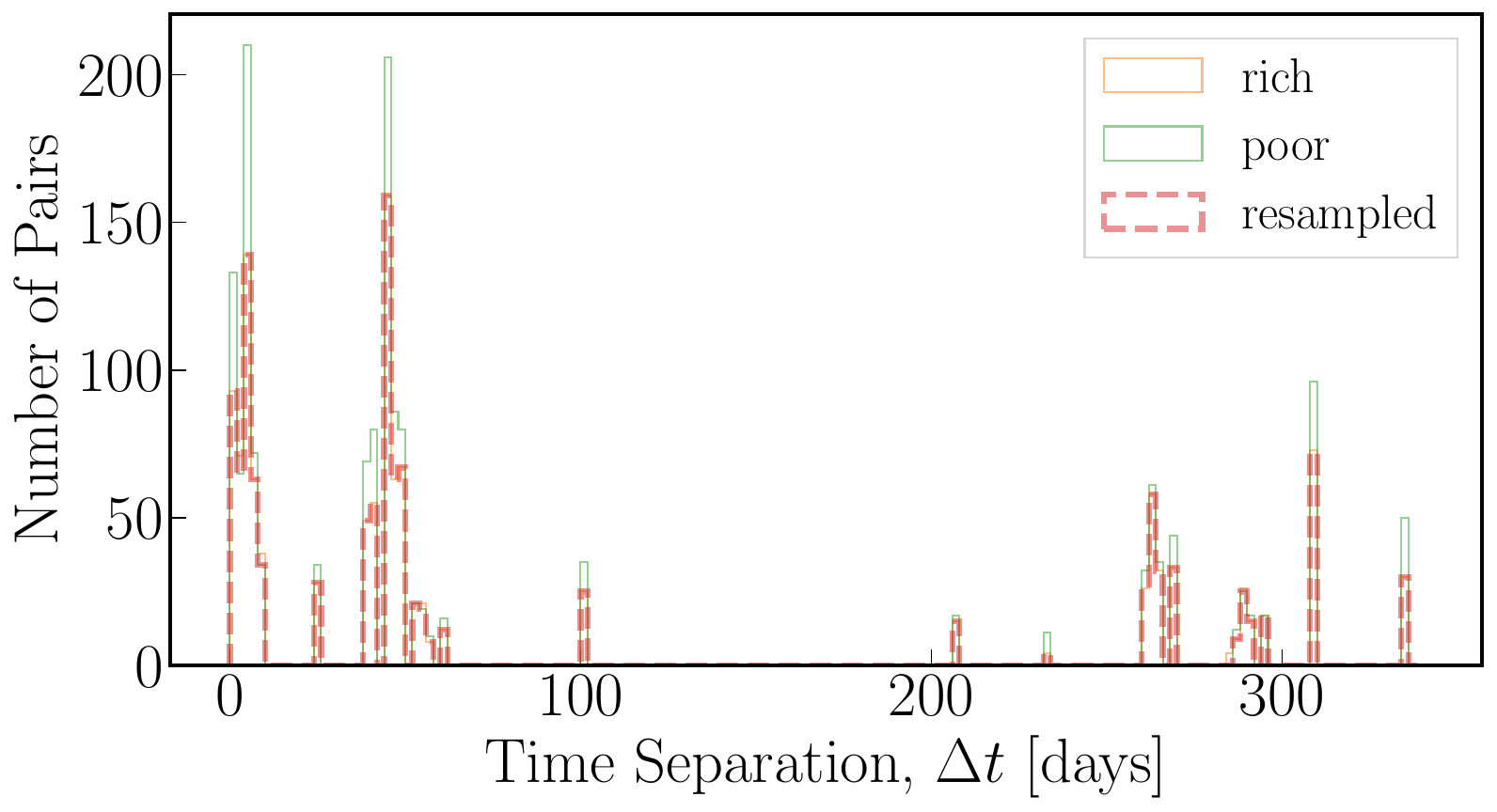}
\plotone{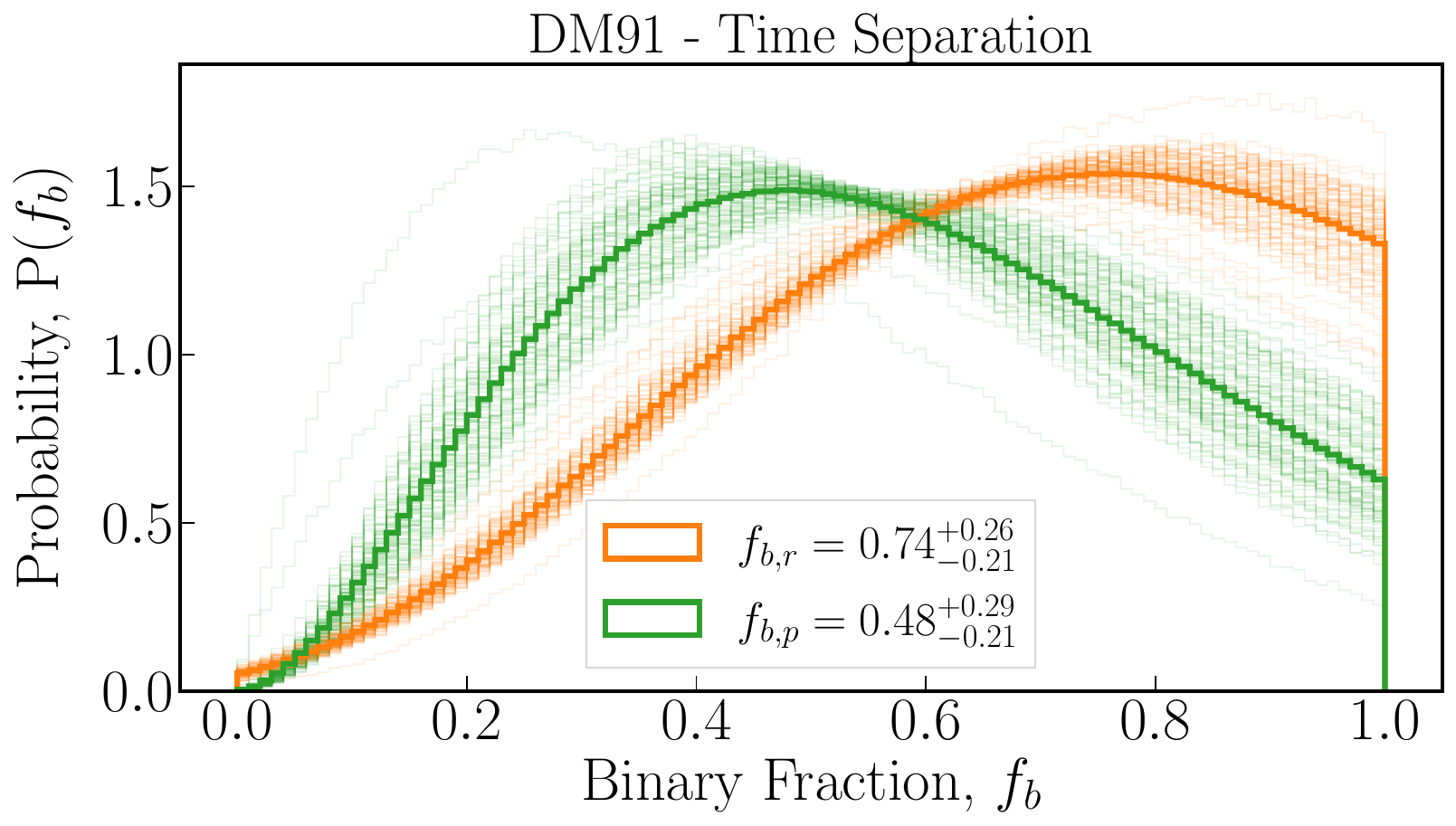}
\plotone{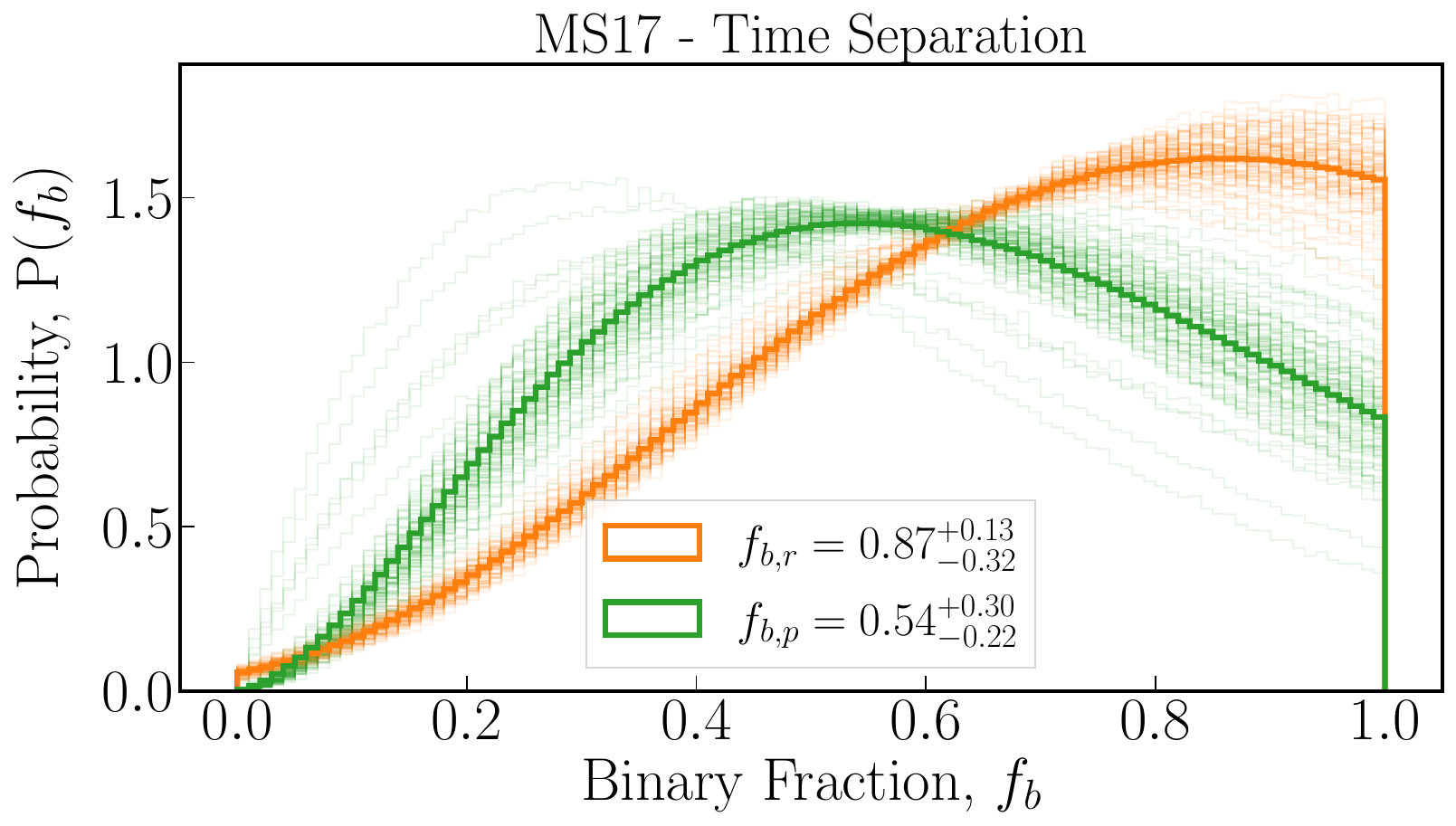}
\caption{
Similar to Figure~\ref{fig:obs_effect}, we resample both the metal-rich and metal-poor subsamples to ensure they share the same distribution of time separations across all observation pairs. 
{\bf Top:} The distribution of the time separation in units of day for individual pairs of observations.
{\bf Middle:} The PPDs derived from the \citetalias{1991A&A...248..485D} model after resampling. The orange histogram is the result of the metal-rich population, while the green one is the result of the metal-poor population.
{\bf Bottom:} The corresponding results from the \citetalias{2017ApJS..230...15M} model.
Both models show consistent results, with the discrepancy between the two subsamples persisting despite the resampling. 
\label{fig:sep_effect}}
\end{figure}

So far, we have resampled the metal-rich and metal-poor populations to make them have the same distribution of number of epochs, radial distance, and LOSV uncertainty distributions. We also want to resample them to have the same distribution of time baseline/separation. 
However, directly throwing away individual stars can hardly ensure the time separation to be the same, because different stars have different observations and pair combinations. 
To address this, we assume that the observation pairs adopted to calculate the LOSV variability ($\beta$, see Equation~\ref{equ:beta}) are independent and resample the pairs to bring the same time separation distributions between the metal-rich and metal-poor populations. 
While this approach is not entirely accurate, it provides a reasonable approximation. 
As shown in Figure~\ref{fig:sep_effect}, after resampling the time separation distribution to be the same for the metal-rich and metal-poor populations, the discrepancy in binary fraction between the two populations remains evident.

Our tests above suggest that the difference in the binary fractions between metal-rich and metal-poor populations cannot be due to observational selection effects. In fact, after resampling the two populations to have the same distributions of number of epochs, radial profile, LOSV uncertainty, and time separation, the significance of the detected binary fraction difference between the two populations becomes slightly higher, which is about 1.5$\sigma$.

\subsection{Systematics Associated with the Period Distribution}
\label{ssec:period}
In addition to observational factors, the inferred binary fraction is highly sensitive to the choice of model parameter distributions. 
As mentioned in Section~\ref{subsec:parameters}, the orbital period distribution plays a major role in constraining the binary fraction, with a lower mean value leading to a substantial decrease in the inferred binary fraction. 

\begin{figure}
\plotone{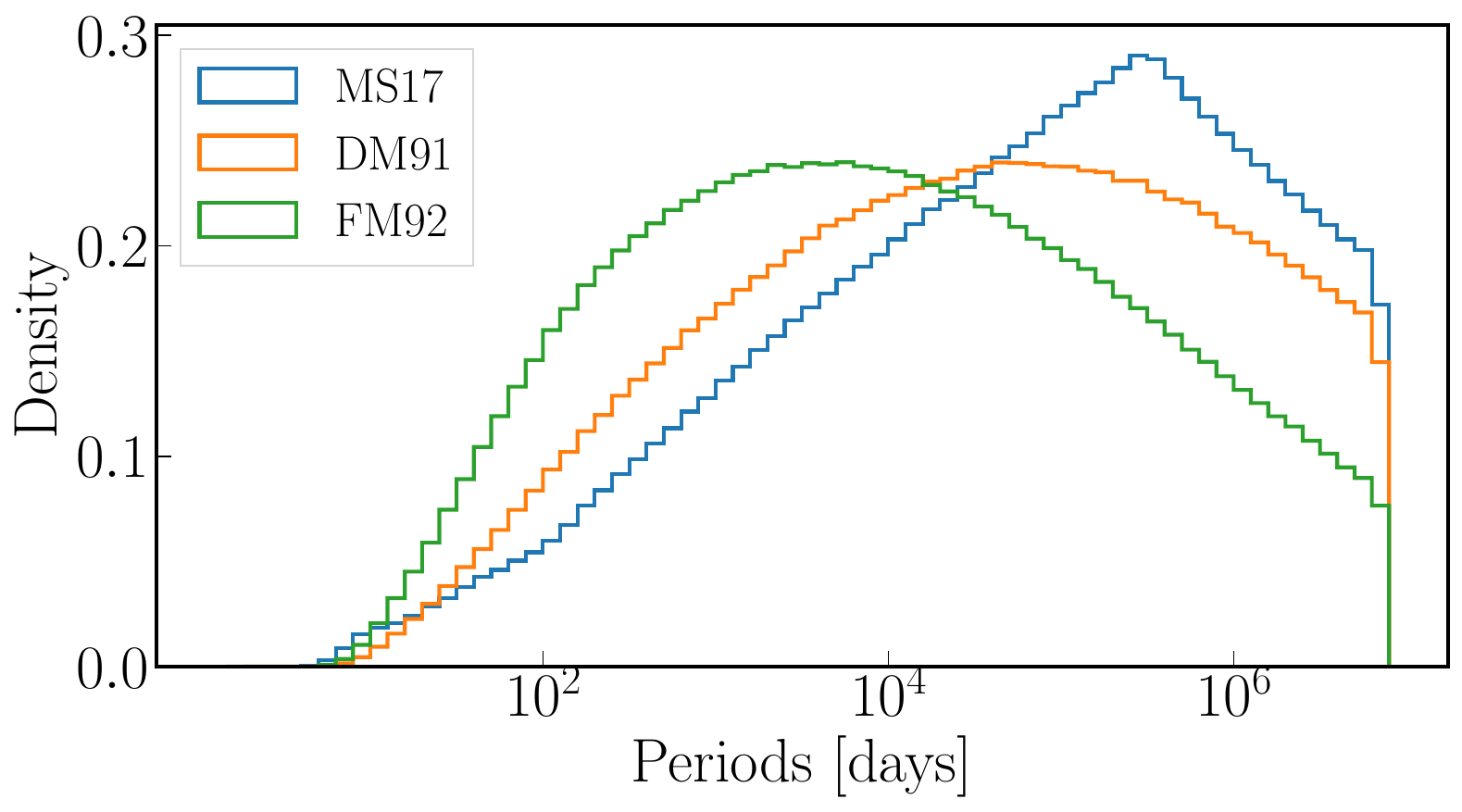}
\caption{The orbital period distributions of the \citetalias{2017ApJS..230...15M}, \citetalias{1991A&A...248..485D}, and \citetalias{1992ApJ...396..178F} models, respectively. 
\label{fig:dist_periods}}
\end{figure}

In Section~\ref{ssec:bf}, the binary fraction derived using the \citetalias{2017ApJS..230...15M} model is slightly higher than that obtained with the \citetalias{1991A&A...248..485D} model. 
The key distinction between these models lies in their treatment of orbital parameter distributions. 
While the \citetalias{1991A&A...248..485D} model assumes independent orbital parameter distributions, the \citetalias{2017ApJS..230...15M} model provides joint distributions that account for correlations between different orbital parameters. 
We present the period distributions in Figure~\ref{fig:dist_periods} for different models. 
Compared to the \citetalias{1991A&A...248..485D} model, represented by the orange line, the \citetalias{2017ApJS..230...15M} model, shown by the blue line, predicts a bit smaller fraction of short-period binaries. This is the major reason that leads to an overall slight increase in the inferred binary fraction by the \citetalias{2017ApJS..230...15M} model, as demonstrated in our results above.

This sensitivity is also demonstrated in \citetalias{2018AJ....156..257S}, which infers that the binary fraction decreases as the mean period value of the model, $\mu_{\log P}$, increases.
\citetalias{2018AJ....156..257S} uses three different models, including \citet{1992ApJ...396..178F} (hereafter \citetalias{1992ApJ...396..178F}), \citetalias{1991A&A...248..485D} and \citet{2011MNRAS.417.1702M}, which differ only in the assumed mean period values of $\mu_{\log P}$ = 3.5, 4.8, and 5.8, respectively.
The corresponding binary fractions derived from these models by \citetalias{2018AJ....156..257S} are 0.52, 0.78, and 0.96, illustrating the strong dependency of the inferred binary fraction on the choice of period distribution.

\begin{figure}
\plotone{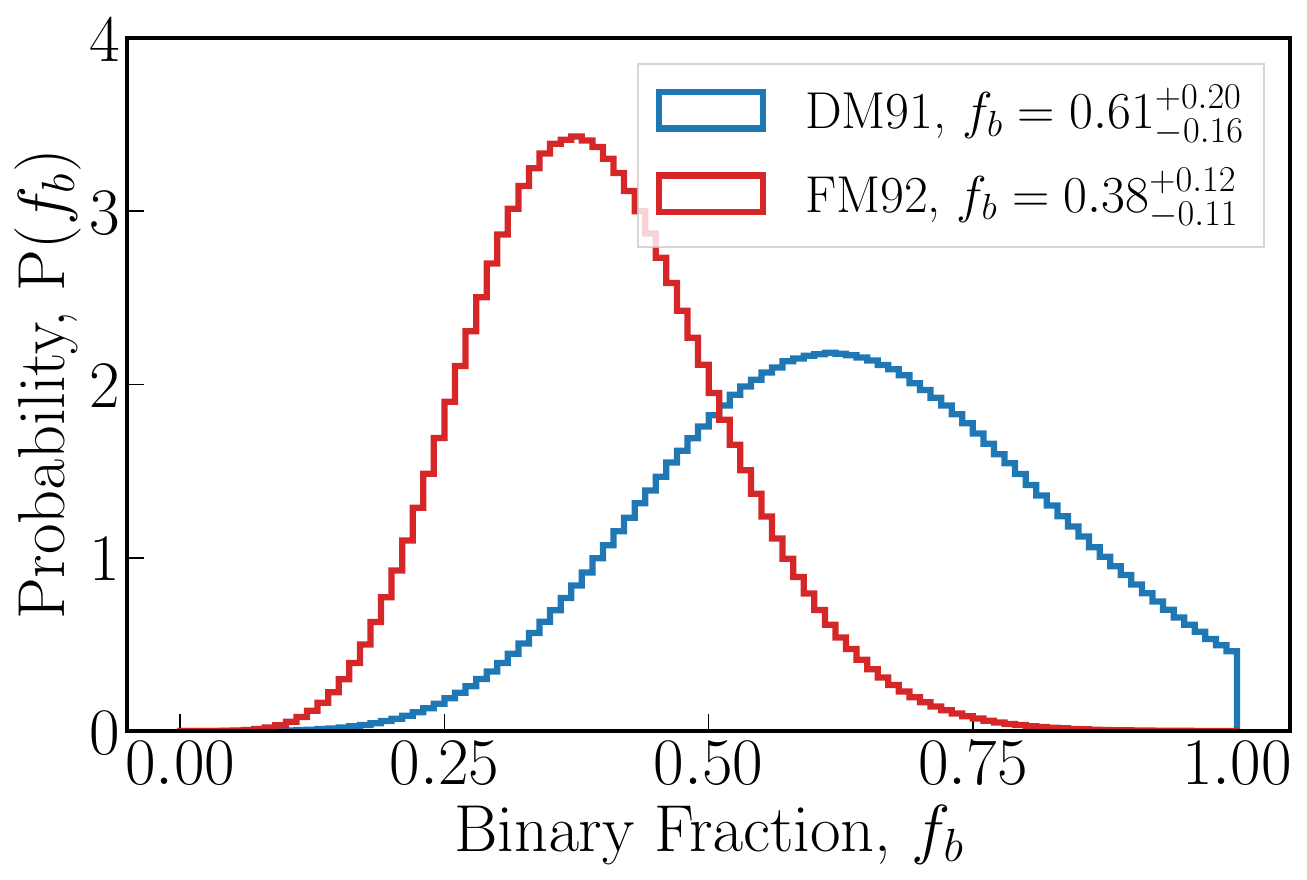}
\plotone{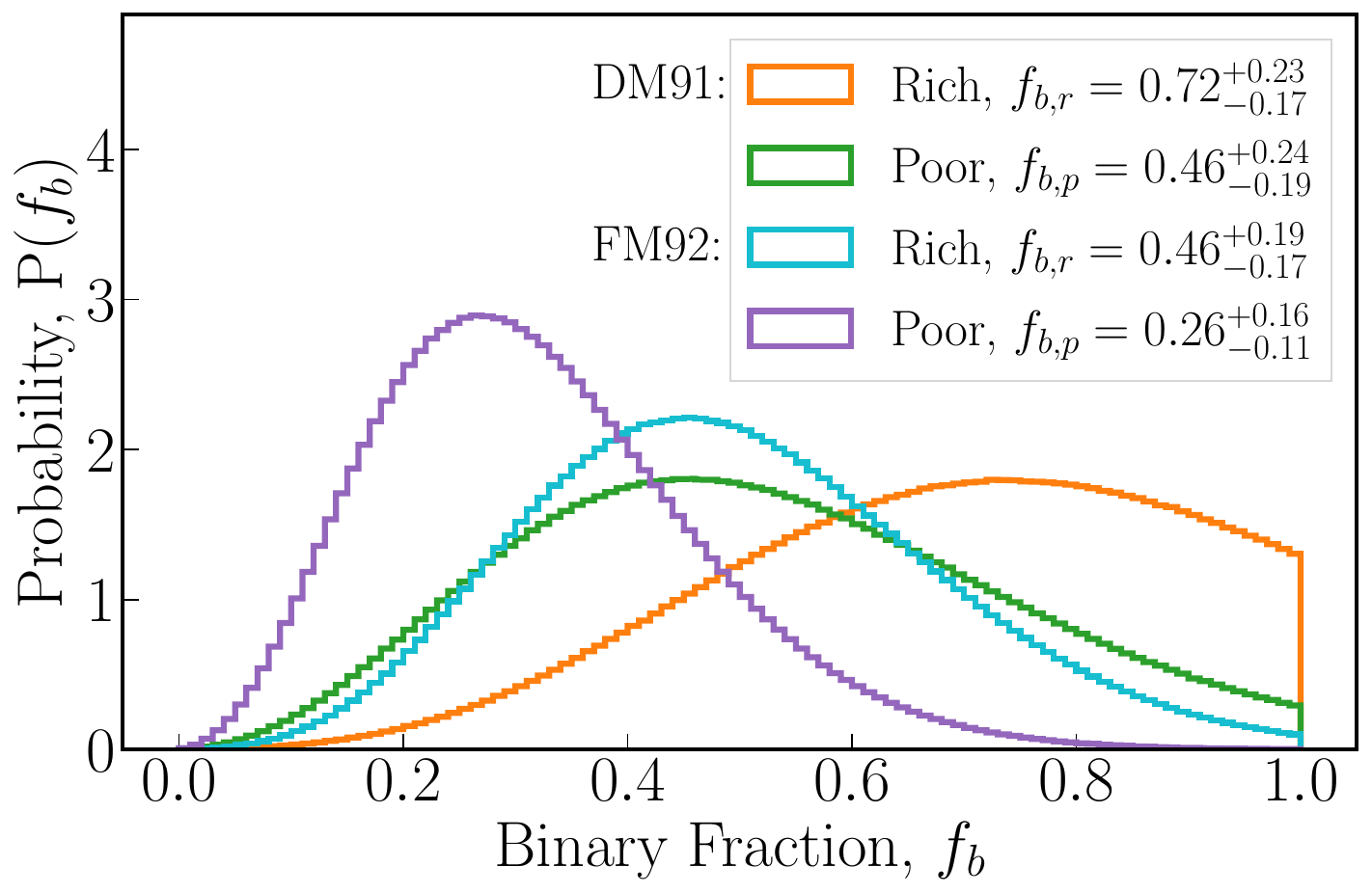}
\caption{{\bf Top:} The PPDs derived from the \citetalias{1991A&A...248..485D} (blue) and \citetalias{1992ApJ...396..178F} (red) models, where the only difference is the shift in the mean value of the log-normal period distribution from 4.8 (\citetalias{1991A&A...248..485D}) to 3.5 (\citetalias{1992ApJ...396..178F}). 
This change causes the binary fraction to decrease significantly, from 0.61 to 0.38, as expected.
{\bf Bottom:} The PPDs for the metal-rich (orange/cyan) and metal-poor (green/purple) subsamples under the \citetalias{1991A&A...248..485D} and \citetalias{1992ApJ...396..178F} models, respectively. 
When a lower mean period is applied, the binary fractions for both subsamples are consistently shifted to lower values.
\label{fig:model_effect}}
\end{figure}

To further investigate the impact of the orbital period distribution, we repeat our constraints adopting the mean period value from the \citetalias{1992ApJ...396..178F} model\footnote{We have also explored the \citet{2011MNRAS.417.1702M} period distribution model, which leads to a binary fraction very close to one and might not be realistic, so we do not show the results based on this model.}, which is lower than the value of  \citetalias{1991A&A...248..485D}, while keeping all other orbital parameter distributions identical to those in \citetalias{1991A&A...248..485D}. 
For clarity, we refer to this modified version as the ``\citetalias{1992ApJ...396..178F}'' model for our discussions afterward.
The green line in Figure~\ref{fig:dist_periods} represents the orbital period distribution for the \citetalias{1992ApJ...396..178F} model, which shows a higher fraction of shorter-period binaries compared to the \citetalias{1991A&A...248..485D} model.
As shown in the upper panel of Figure~\ref{fig:model_effect}, the binary fraction derived from the \citetalias{1992ApJ...396..178F} model decreases by approximately 0.2 for the full sample compared with that from \citetalias{1991A&A...248..485D}. This difference arises primarily due to the shift towards shorter orbital periods, and thus a lower binary fraction is preferred for the same LOSV variability. 

The bottom panel of Figure~\ref{fig:model_effect} shows that the binary fractions for both the metal-rich and metal-poor subsamples decrease by approximately the same amount when applying the \citetalias{1992ApJ...396..178F} model with a lower mean period. Despite the decrease in the inferred binary fraction, the discrepancy between the metal-rich and metal-poor populations persists (see the cyan and purple curves). Here we emphasize that we are applying the same orbital period distribution model to both metal-rich and metal-poor populations. If the metal-rich and metal-poor populations follow different period distributions, such as a shorter mean period for the metal-rich population and a longer mean period for the metal-poor population, the observed discrepancy in binary fractions could be reduced or even eliminated. This effect can be seen by comparing the green curve of the metal-poor population and the cyan curve for the metal-rich population in the bottom panel of Figure~\ref{fig:model_effect}.
However, explaining the discrepancy in this way would require the metal-rich and metal-poor populations to have fundamentally different orbital period distributions, for which there is no strong theoretical support and which is therefore unlikely to be physically realistic.

\subsection{Is one year of time baseline enough?}
\label{ssec:baseline}

\begin{figure}
\plotone{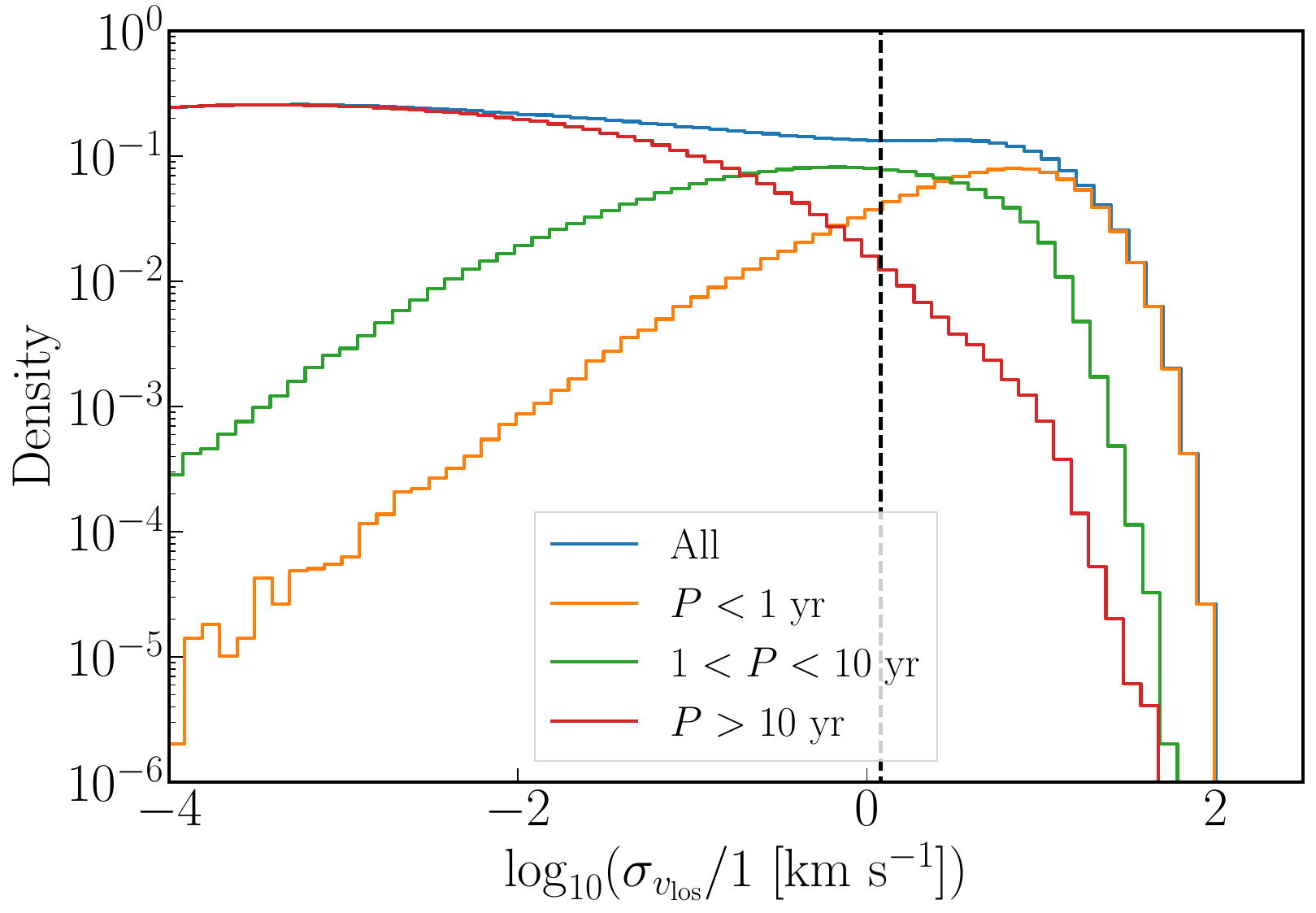}
\caption{LOSV variability for simulated binary systems with different orbital periods. A binary fraction of 100\% is adopted for the simulation. The variability is characterized by the standard deviation of the LOSV of each object ($\sigma_{v_{\rm{los}}}$).
Here unlike the simulations that we have done for the main results of the paper, we choose not to include any additional noise. The blue histogram represents the $\sigma_{v_{\rm{los}}}$ distribution for the entire simulated sample, while the other histograms correspond to a few different ranges of orbital period, as indicated by the legend. The black dashed line at $\sigma_{v_{\rm{los}}}=1.2\ \mathrm{km\ s}^{-1}$ denotes the detection threshold, which is the systematic error floor of LOSV measurements from the member stars in UMi. Objects with LOSV variability below this level do not contribute to the inferred probability in either the observational data or the simulations. 
\label{fig:diff_period}}
\end{figure}

Given the sensitivity of the binary fraction to the period distribution, as we have discussed in the previous subsection, it could be a concern whether the limited time baseline of one year for the DESI UMi observation is sufficient to accurately recover the binary fraction. 
The one-year coverage may not capture velocity variations of longer-period binaries, potentially biasing our results towards short-period binary systems. Therefore, it is crucial to assess whether our methodology can reliably constrain the binary fraction with one year of time baseline.
In this section, we perform additional tests to examine the reliability of our methodology through mock simulations. 

We first assess the fraction of binary systems that can be detected with the one-year observational baseline. 
To achieve this, we generate a mock sample of binaries by randomly selecting 1,000 samples from our simulations (see Section~\ref{ssec:MCS}), which contains about 670,000 stars with over 2 million exposures in total. The large sample size ensures that statistical fluctuations are small enough.
In this sample, the binary fraction is set to 1. Besides, we do not include any Gaussian noise, which is described in step 4 of Section~\ref{ssec:MCS}, to isolate the intrinsic LOSV variations purely induced by the orbital motion of the binaries. 

We use the standard deviation of the LOSV, $\sigma_{v_{\rm{los}}}$, to quantify velocity variations in this test.
The results from the mock binary sample are shown in Figure~\ref{fig:diff_period}, where the distribution of $\sigma_{v_{\rm{los}}}$ is represented by the blue histogram. 
As illustrated in the figure, approximately 12\% of the simulated binaries exhibit LOSV variations exceeding 1.2~ km s$^{-1}$, as indicated by the black vertical dashed line. This threshold represents the systematic error floor of LOSV measurements of the member stars in UMi from the \textsc{rvs} pipeline, as mentioned in Section~\ref{sec:data}. 
Due to the limited one-year observational baseline, the LOSV variations of long-period binaries remain small. Examining the three additional histograms, which divide the sample based on orbital periods, we observe a clear trend. The majority ($99.6\%$) of long-period binaries ($P>10$ years, red histogram) exhibit velocity variations of $\sigma_{v_{\rm{los}}} < 1.2$ km $s^{-1}$, making them difficult to detect. 
In contrast, short-period binaries ($P<1$ year, orange histogram) show significantly larger velocity variations, while the intermediate-period binaries ($1<P<10$ years, green histogram) fall between these two groups, displaying moderate velocity variations.
Among the entire binary sample, about 12\% of stars with LOSV variations surpassing the 1.2~km s$^{-1}$ error floor provide meaningful constraints on the binary fraction. 
The remaining stars, with LOSV variations below this threshold, are dominated by uncertainties and contribute negligibly to our analysis.

The analysis above reveals that binaries with periods longer than $P\simeq10$ years are more difficult to detect with a one-year time baseline, and thus our constraints on the overall binary fraction largely rely on extrapolating the LOSV variabilities revealed by short-period binaries. 
To check whether the extrapolation is robust, we conduct additional tests to determine whether the limited subset of stars with detectable LOSV variations is adequate for a robust binary fraction estimation. Specifically, we generate mock datasets with predefined binary fractions and introduce realistic observational uncertainties consistent with those in our actual data. 
We then apply exactly the same methodology used in our main analysis, treating the true binary fraction as unknown. 
By comparing the recovered binary fractions to the input values, we assess the robustness of our method and evaluate whether the 12\% stars within the limited observational time window allow for reliable recovery of the binary fraction.

\begin{figure}
\plotone{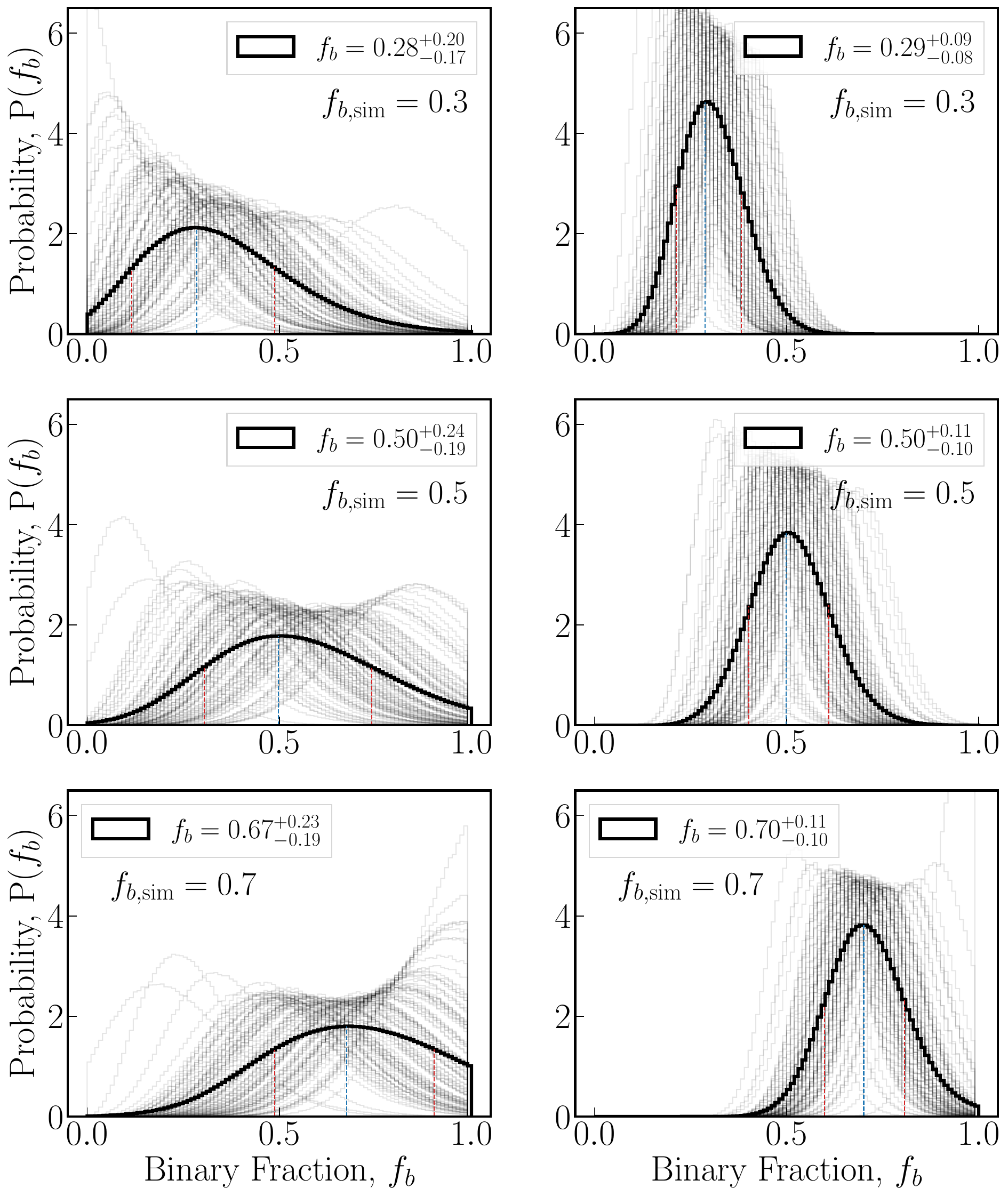}
\caption{PPDs of the binary fraction derived from six mock datasets with input binary fractions of 0.3, 0.5, and 0.7, shown in the top, middle, and bottom panels, respectively. All PPDs are computed using the same method as in our main analysis.
Black solid lines represent the combined result from all 100 mock subsamples, and the grey shaded lines correspond to individual subsample realizations.
Blue dashed lines indicate the peak of the posterior distribution, while red dashed lines mark the $1\sigma$ credible intervals.
The left column shows results from mock datasets matched in size to the real sample, while the right column presents results from mock samples with five times more stars, demonstrating the improved precision with increased sample size.
}
\label{fig:simu_bf}
\end{figure}

In Figure~\ref{fig:simu_bf}, we show the PPDs derived from three mock datasets, each initialized with a binary fraction of 0.3, 0.5, and 0.7, respectively. 
For each case, we generate two sets of mock samples: one with the same number of member stars as our real dataset, and another with five times more stars to evaluate the impact of sample size on the precision of our estimates.
The results from the equal-size samples are shown as gray dashed histograms, while those from the larger samples are plotted as solid dark histograms. 
In all cases, the inferred binary fractions are consistent with the input values, demonstrating the accuracy and reliability of our method. 
As expected, the uncertainties are significantly reduced when using larger sample sizes.

Our analysis thus demonstrates that the one-year observation baseline is sufficient to recover the overall binary fraction, though it relies on extrapolations of LOSV variabilities for short-period binaries. 
Long-period binaries, which exhibit smaller LOSV variations over short timescales, remain unresolved within this limited timescale. 
Despite this limitation, our estimated binary fraction is not expected to be underestimated.  Even without directly detecting long-period binaries, our model allows us to extrapolate the total contribution based on the observed short-period binaries.

In our test associated with Figure~\ref{fig:simu_bf}, however, we adopted exactly the same binary orbital parameter distribution model used for generating the mock data to recover the binary fraction. 
So the model extrapolation is perfect. 
If, in reality, the underlying binary orbital parameter distributions differ from what we assume of using the \citetalias{1991A&A...248..485D} or \citetalias{2017ApJS..230...15M} models, the extrapolation may not work as perfectly as what we have demonstrated with Figure~\ref{fig:simu_bf} to recover the actual binary fractions. 
Nevertheless, we can at least conclude that for short-period binaries that can be detected within a one-year time baseline, the metal-rich population has slightly higher binary fractions than the metal-poor population. Future observations, including more dedicated tiles of DESI MWS, can further increase the time baselines and will be crucial for obtaining more comprehensive and precise constraints on the binary fractions for longer period binaries in UMi, in other dwarf galaxies, in globular clusters, and in the field of our MW.

In addition, these simulations allow us to directly assess the impact of binaries on the measured velocity dispersion. By analyzing the mock binary sample described above and assuming a binary fraction of 1, we derive the velocity dispersion induced solely by binary orbital motions. As we mentioned above, long-period binaries ($P>10$ years) produce velocity variations that are indistinguishable from measurement uncertainties and therefore contribute negligibly to the likelihood. Consistently, their impact on the inferred velocity dispersion is also minimal, at the level of less than 1\%, which does not affect our subsequent analysis. 
Consequently, even in the extreme case of a high intrinsic binary fraction (assumed to be 1), long-period binaries do not inflate the total velocity dispersion; instead, the dominant contribution arises from short- and intermediate-period systems, which could increase the dispersion by approximately 17\% if all binaries are assumed to have periods shorter than 10 years. When binaries of all periods are included, the overall inflation is about 4\%.

\section{Discussions}
\label{sec:discussion}

Our analysis reveals that the metal-poor stellar population in UMi exhibits a slightly lower binary fraction compared to the metal-rich population. 
Given the results of the selection effect analysis in Section~\ref{ssec:selecteff} and the discussion in Section~\ref{ssec:period}, we can rule out observational biases as the primary drivers of this detection. 
To better understand these results, it is useful to compare them with previous studies that have explored binary fraction trends across different stellar populations.

Several studies have investigated how binary fractions vary with stellar properties such as mass, metallicity, and age. 
For instance, \citet{2010ApJS..190....1R} conducts a comprehensive investigation of companions to solar-type stars up to 25~pc, showing that more massive, bluer stars are more likely to host binary companions than redder, lower-mass stars. 
Additionally, they find that among redder stars, metal-poor systems exhibit a higher probability of hosting companions. 
Similarly, \citet{2018RAA....18...52T} examines the relationship between binary fraction and $\alpha$-element abundances. 
They find that older stars with higher $[\alpha/\mathrm{Fe}]$ tend to have a higher binary fraction. \citet{2018ApJ...854..147B} investigated the multiplicity of field stars. 
They also reveal a metallicity dependency, with metal-poor stars ($[\mathrm{Fe/H}] \lesssim -0.5$) exhibiting a multiplicity that is 2–3 times higher than that of metal-rich stars ($[\mathrm{Fe/H}] \gtrsim 0.0$). Besides, the study by \citet{2019ApJ...875...61M} finds that the close binary fraction ($P<10^4$ days; $a<10$ au) among solar-type stars decreases as the metallicity increases. 
Another work by \citet{2019MNRAS.484.2341B} reports a similar trend from a few radiation hydrodynamical simulations, that the multiplicity slightly increases as metallicity decreases, which is due to the greater cooling rates at high gas densities due to the lower opacities at low metallicities increasing the gas fragmentation on small scales.
These studies all suggest that stars formed earlier, which are typically more metal-poor and $\alpha$-enhanced, have higher binary fractions.

It seems our analysis reveals a slightly opposite trend from those previous studies: stars in UMi with lower [Fe/H] and thus older, have smaller binary fractions. However, direct comparisons between our results and previous studies are not straightforward. 
The metal-poor populations in these previous works are not sufficiently metal-poor to match the conditions of UMi. As shown in the lower left panel of Figure~\ref{fig:info}, even the most metal-rich member star in our UMi sample has $\mathrm{[Fe/H]}\simeq-1$, reflecting the extremely low metallicity environment in UMi. Our analysis focuses on a regime that is significantly more metal-poor than those in earlier studies.
As a comparison, the most metal-poor stars in \citet{2018RAA....18...52T} have $\mathrm{[Fe/H]}>-1.0$, a value that would be considered part of the metal-rich population in our analysis. 
Although the stars analyzed in \citet{2018ApJ...854..147B} span a broad metallicity range down to $[\mathrm{Fe/H}] \approx -2.5$, the majority is from $-$1 to 0.5, and their metal-rich sample is defined as $[\mathrm{Fe/H}] \gtrsim 0.0$. 
Similarly, Figure~15 of \citet{2019MNRAS.484.2341B} shows that their sample only covers metallicities greater than 0.01 $\rm{Z}_\odot$.
In the case of \citet{2010ApJS..190....1R}, they only include stars with $\mathrm{[Fe/H]}>-0.9$, again significantly more metal-rich than our case. Furthermore, most studies primarily focus on solar-type stars around our neighborhood, where the average metallicity is already enhanced compared to that of dwarf galaxies. Given these differences, we cannot directly apply their conclusions to our study. 

A crucial factor to consider might be the significantly older age of the stellar population in our sample than those studied in previous works.
Dynamical evolution likely plays a role in shaping the observed binary fraction. Older, metal-poor stars have undergone a longer time for dynamical interactions to happen, increasing the risk of disruption due to encounters with flyby stars or other tidal interactions. 
As suggested by previous works \citep{2010arXiv1005.5388P,2016MNRAS.461L..72P,2022ApJ...929...77K}, wide stellar binaries in dwarf spheroidal galaxies can be disrupted by repeated encounters with dark substructures orbiting within their dark matter halos. 
The much more extended evolutionary history for older stars formed in dwarf galaxies may explain the lower binary fraction observed in metal-poor populations. For the other previous studies mentioned above, their sample of stars may not have undergone evolutionary histories that are as old as member stars in old dwarf galaxies, so they may not have enough time to accumulate enough dynamical encounters to disrupt binary systems.

\begin{figure}
\plotone{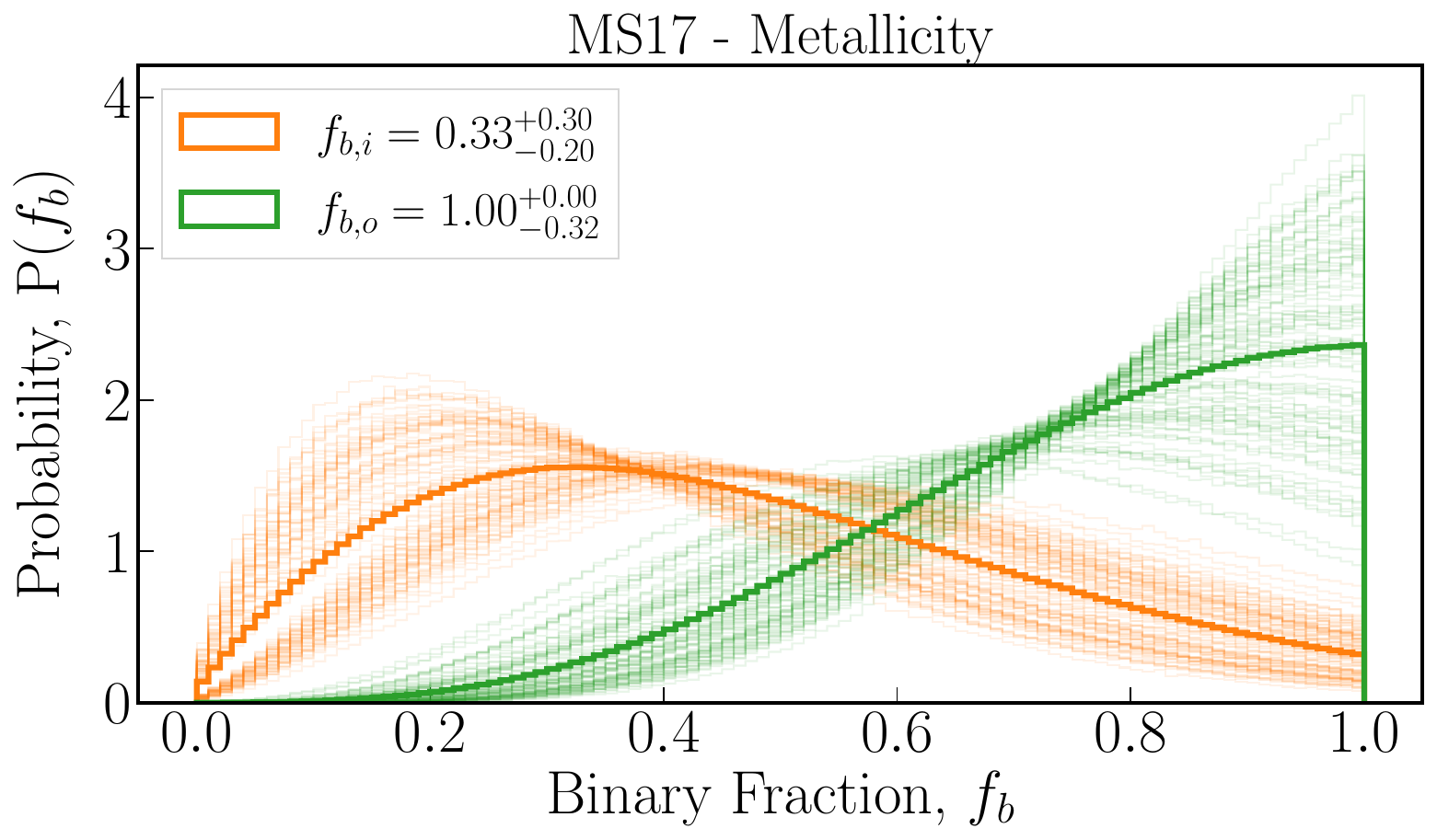}
\caption{PPDs of the binary fraction for the inner (orange) and outer (green) subsamples, after resampling to match the same metallicity distribution, divided according to the projected semi-major axis of 0.92 degrees (see the red ellipse in Figure~\ref{fig:info}), corresponding to the elliptical isophotes of the number density. }
\label{fig:inout}
\end{figure}

If the explanation of long-time dynamical interactions to disrupt binaries in the metal-poor population holds, we may expect the central part of UMi, which is denser, to have more frequent encounters and lower binary fractions. We test this in Figure~\ref{fig:inout}, in which we divide the member stars in UMi into inner and outer subsamples, according to the projected semi-major axis of 0.92 degrees, which is marked by the red ellipse in Figure~\ref{fig:info}. Since the inner part is more metal-rich, we resample both populations to have the same metallicity distribution following the approach in Section~\ref{ssec:selecteff}. In this way, we do find that the inner part has a lower binary fraction\footnote{The result here indicates that for our main conclusion about the metallicity dependency of the binary fraction, it is important to control the metal-rich and metal-poor populations to have the same radial distribution (middle column of Figure~\ref{fig:obs_effect}). On the other hand, the change in the best inferred binary fractions after resampling the number of epochs and time baselines of different populations is actually mainly contributed by the radial dependency of binary fractions. In fact, we have tested with mock data that as long as the observed stars and our associated Monte Carlo simulations have matched distributions of number of epochs and time baselines (see step 2 of Section~\ref{ssec:MCS}), the inference is not biased even if we choose to randomly discard some epochs.} of $0.33^{+0.30}_{-0.20}$, whereas stars in more outskirts have a higher binary fraction of $1.00^{+0.00}_{-0.32}$. The significance is slightly smaller than 2$\sigma$. This seems to be consistent with our hypothesis. 

However, as we have mentioned in Section~\ref{ssec:baseline} above, our one-year time baseline makes it hard to detect the LOSV variability of binaries with orbital periods longer than 10 years, for which the best-constrained binary fractions rely on model extrapolations. For binaries with orbital periods of 1 to 10 years that can be more safely detected by our time window, the typical binary separation is at most a few AU, assuming a circular orbit and a mass ratio of 1. However, given the typical density of stars in MW dwarf galaxies, we estimate the typical mean free path of close encounters to be 1000~AU in UMi, assuming the same stellar number density as in Section~\ref{sssec:P}. This suggests that the separations of short-period binaries that can be detected by our observations are too small to be sensitive to close encounters. Thus, the explanation of long-time dynamical interactions to disrupt binaries in the metal-poor population of dwarf galaxies may only be valid if our model extrapolations hold for longer-period binaries. 

Another explanation associated with the disruption of short-period binaries may be related to the orbital decay and eventual merger driven by dynamical friction in their early evolutionary phase. When young binaries form within dense molecular clouds, they experience drag from the surrounding gas, which carries away angular momentum through acoustic disturbances in the form of spiral waves \citep[e.g.][]{2010MNRAS.402.1758S}.

One more alternative explanation lies in the conditions of the early galaxy environment. During the initial stages of galaxy formation, the interstellar medium was hotter and more turbulent. The metal-poor population likely formed in much denser regions, where frequent tidal interactions made it difficult for binaries to survive. In contrast, the metal-rich population formed later in a more diffuse environment with a less dense interstellar medium, allowing a greater fraction of binaries to remain intact. This difference in formation conditions may lead to the observed disparity in the binary fraction between the two populations. 

In the end, we mention recent studies by \citet{2025AJ....170..171L} and \citet{2025A&A...702A..11G}, which provide independent evidence that is more consistent with our findings. \citet{2025AJ....170..171L} identified nearly equal-mass binaries through the difference between the geometric absolute magnitudes from {\it Gaia} parallax and the absolute magnitudes inferred from spectrophotometric distances measured by a data-driven approach. Their results indicate a positive correlation between binary fraction and metallicity, consistent with our findings in UMi. Moreover, a recent analysis of BHB stars by \citet{2025A&A...702A..11G} found that halo-like, metal-poor populations exhibit lower intrinsic binary fractions than their more metal-rich, disk-like counterparts. 
This contrast was interpreted as evidence that halo BHBs likely formed predominantly through single-star evolutionary channels, while disk BHBs experienced more binary-driven evolution. Their finding aligns broadly with our result. 

The influence of metallicity on binary fractions in metal-poor environments still remains largely unexplored. Our findings provide possible new probes into how binary populations evolve under such conditions and how long-term dynamical processes, combined with early formation environments, may shape their survival over time. However, the significance of our detection is marginal, and we need future observations with longer time baselines to refine confirm these trends, as well as to further understand the fundamental mechanisms governing binary evolution in dwarf galaxies.

\section{Conclusion}
\label{sec:conclusion}

In this paper, we conduct a detailed investigation of the binary fraction in the Ursa Minor dwarf galaxy using spectroscopic data from the DESI Milky Way Survey. 
Our dataset consists of multiple exposures taken over approximately one year, providing a large sample of LOSV measurements that enable us to track LOSV variability as a proxy for binary abundance.

Adopting the orbital parameters from \citet{1991A&A...248..485D} (hereafter \citetalias{1991A&A...248..485D}) and \citet{2017ApJS..230...15M} (hereafter \citetalias{2017ApJS..230...15M}) models,
we derive overall binary fractions for UMi of $0.61^{+0.16}_{-0.20}$ and $0.69^{+0.19}_{-0.17}$, respectively. 
These estimates are consistent with previous studies, demonstrating that high binary fractions are present in UMi.

After dividing member stars in UMi into two approximately equal-sized subsamples by metallicity, we find a discrepancy between the inferred binary fractions of the metal-rich and metal-poor populations, with metal-rich stars exhibiting slightly higher binary fractions. In the \citetalias{1991A&A...248..485D} model, metal-rich sample reaches $0.72^{+0.23}_{-0.17}$, while metal-poor sample drops to $0.46^{+0.24}_{-0.19}$. Similarly, with the \citetalias{2017ApJS..230...15M} model, these values are $0.86^{+0.14}_{-0.24}$ and $0.48^{+0.26}_{-0.19}$ for the metal-rich and metal-poor subsamples, respectively.

Metal-rich and metal-poor stars in UMi have different spatial distributions, with metal-rich stars more centrally concentrated. The observing strategy of UMi involves more tiling in central regions and, hence, may result in slightly different distributions of number of epochs, time separations, and LOSV measurement uncertainties for metal-rich and metal-poor populations. We employ a resampling approach to mitigate such observational biases. By ensuring that both subsamples share comparable distributions of observational parameters after resampling, the difference between the binary fractions in the metal-rich and metal-poor populations persists, indicating the difference is not due to selection effects. However, our time window of observation for UMi is only 1~year, which makes it difficult to capture the velocity variability for binaries with orbital periods longer than 10~years, and thus our conclusion about longer-period binaries strongly depends on model extrapolations.

The lower binary fraction in metal-poor stars, if genuine, may be linked to a combination of factors, including dynamical evolution and the conditions of early star formation. Metal-poor stars, which formed in a denser and possibly more turbulent interstellar medium at earlier times, may have experienced more frequent binary disruptions due to, for example, close stellar encounters, compared to metal-rich stars, which formed in a later, more chemically evolved, and dynamically stable environment, allowing more binaries to survive. Additionally, metal-poor stars in UMi are older on average, and thus there would have been a longer time for the effect of stellar encounters to accumulate and finally disrupt the binary system than those in the field. 

We also divide member stars in UMi according to the projected radius to the center, and after resampling them to have the same metallicity distributions, we find that stars in the denser central region have lower binary fractions ($0.33^{+0.30}_{-0.20}$) than the outer part ($1.00^{+0.00}_{-0.32}$) with the \citetalias{2017ApJS..230...15M} model. 

Future surveys with more exposures crossing longer time baselines and higher signal-to-noise spectroscopy will be crucial in further verifying these findings. 
Although the present analysis achieves high coverage of confirmed member candidates, matching over 90\% of the current target list, only about 50\% of these stars are included in the final analysis due to insufficient epoch coverage or large LOSV uncertainties.
Extended observations that provide additional epochs and improved velocity precision will therefore enhance sensitivity to longer-period binaries and enable more robust constraints on the binary population in UMi. 
With improved sampling of orbital periods and tighter constraints on the binary fraction, future studies will be better positioned to explore the physical mechanisms governing binary survival in metal-poor environments.
Moreover, similar studies in other dwarf galaxies within the Milky Way will help to determine whether the metallicity trends observed in UMi are applicable more broadly across different environments. 
This will ultimately deepen our understanding of binary star evolution, especially in the context of early galaxy formation and the complex interplay among chemical evolution, star formation, and dynamical interaction. 

The results presented in this paper can be accessed at \url{https://doi.org/10.5281/zenodo.16827505}, which contains all data points for the figures.

\begin{acknowledgments}

This work is supported by NSFC (12573022, 12595312, 12273021), the National Key R\&D Program of China (2023YFA1605600, 2023YFA1605601), 111 project (No.\ B20019), and the Office of Science and Technology, Shanghai Municipal Government (grant Nos. 24DX1400100, ZJ2023-ZD-001). We thank the sponsorship from Yangyang Development Fund. The computations of this work are carried on the National Energy Research Scientific Computing Center (NERSC) and the Gravity supercomputer at the Department of Astronomy, Shanghai Jiao Tong University. SK acknowledges support from the Science \& Technology Facilities Council (STFC) grant ST/Y001001/1. 

This material is based upon work supported by the U.S. Department of Energy (DOE), Office of Science, Office of High-Energy Physics, under Contract No. DE–AC02–05CH11231, and by the National Energy Research Scientific Computing Center, a DOE Office of Science User Facility under the same contract. Additional support for DESI was provided by the U.S. National Science Foundation (NSF), Division of Astronomical Sciences under Contract No. AST-0950945 to the NSF’s National Optical-Infrared Astronomy Research Laboratory; the Science and Technology Facilities Council of the United Kingdom; the Gordon and Betty Moore Foundation; the Heising-Simons Foundation; the French Alternative Energies and Atomic Energy Commission (CEA); the National Council of Humanities, Science and Technology of Mexico (CONAHCYT); the Ministry of Science, Innovation and Universities of Spain (MICIU/AEI/10.13039/501100011033), and by the DESI Member Institutions: \url{https://www.desi.lbl.gov/collaborating-institutions}. Any opinions, findings, and conclusions or recommendations expressed in this material are those of the author(s) and do not necessarily reflect the views of the U. S. National Science Foundation, the U. S. Department of Energy, or any of the listed funding agencies.

The authors are honored to be permitted to conduct scientific research on I'oligam Du'ag (Kitt Peak), a mountain with particular significance to the Tohono O’odham Nation.

We thank all contributors to this work for their valuable input and support. We thank the DESI operation team, Stephen Bailey, Anand Raichoor, and Edward Schlafly for their contributions to the maintenance of the tertiary programs in the DESI MWS. We thank our DESI publication handlers, Ting Tan and Alejandro Avilés, for their assistance. We are grateful to the organizers and participants of the International Astronomical Union (IAU) Symposium 398 for the insightful discussions and constructive feedback that helped shape and improve this study.

For the purpose of open access, the author has applied a Creative Commons Attribution (CC BY) licence to any Author Accepted Manuscript version arising from this submission. 

\end{acknowledgments}

\bibliography{sample631}{}
\bibliographystyle{aasjournal}



\end{document}